\documentclass[12pt,twoside]{article}
\usepackage{overcite}
\usepackage{epsfig}
  \usepackage{amsthm,amssymb}
\usepackage{amsbsy,amsfonts,amsmath}
\makeatletter 
\def\@cite#1{\mbox{$\m@th^{(\hbox{\@ove@rcfont#1})}$}}
\renewcommand\@biblabel[1]{{#1}.}

\def\@sect#1#2#3#4#5#6[#7]#8{\ifnum #2>\c@secnumdepth \def\@svsec{}\else
 \refstepcounter{#1}\edef\@svsec{\csname the#1\endcsname.\ }\fi
 \@tempskipa #5\relax \ifdim \@tempskipa>\z@ \begingroup #6\relax
 \@hangfrom{\hskip #3\relax\@svsec}{\interlinepenalty \@M #8\par} \endgroup
 \csname #1mark\endcsname{#7}\addcontentsline
 {toc}{#1}{\ifnum #2>\c@secnumdepth \else
 \protect\numberline{\csname the#1\endcsname}\fi #7}\else
 \def\@svsechd{#6\hskip #3\relax \@svsec #8\csname #1mark\endcsname
 {#7}\addcontentsline {toc}{#1}{\ifnum #2>\c@secnumdepth \else
 \protect\numberline{\csname the#1\endcsname}\fi #7}}\fi \@xsect{#5}}
\def\section{\@startsection {section}{1}{\z@}{-3.5ex plus -1ex minus
 -.2ex}{2.3ex plus .2ex}{\sffamily\bfseries\normalsize}}
\def\subsection{\@startsection{subsection}{2}{\z@}{-3.25ex plus -1ex minus
 -.2ex}{1.5ex plus .2ex}{\sffamily\bfseries\normalsize}}
\def\subsubsection{\@startsection{subsubsection}{3}{\z@}{-3.25ex plus
 -1ex minus -.2ex}{1.5ex plus .2ex}{\sffamily\bfseries\normalsize}}
\long\def\@makecaption#1#2{%
  \vskip\abovecaptionskip
  \sbox\@tempboxa{\sffamily\bfseries{#1. #2}}%
  \ifdim \wd\@tempboxa >\hsize
   \footnotesize #1.\ \ #2\par
  \else
    \hbox to\hsize{\hfil\box\@tempboxa\hfil}%
  \fi
  \vskip\belowcaptionskip}

\def\footnoterule{\kern-3\p@
  \hrule \@width 0.5in \kern 2.6\p@}
\renewcommand\@makefntext[1]{%
    \parindent 0em%
    \noindent
    \hb@xt@.5em{\hss\@makefnmark}\hskip.2em#1}
 \makeatother

\font\mycal=eusb10 at 12pt
\def\cm{{\mycal{r}}}
\def\cn{\mycal{s}}
\def\hg{\tilde{g}}
\def\bg{\bar{g}}
\def\caption#1{{\footnotesize #1}\smallskip}
\def\cases#1{\left\{#1\right.}
\begin{document}
\pagestyle{myheadings}
\makeatletter
\renewcommand\@evenhead{\sffamily\bfseries\footnotesize\thepage
\hfil Au-Yang and Perk}
\renewcommand\@oddhead{\sffamily\bfseries\footnotesize
$\boldsymbol{Q}$-Dependent Susceptibilities
of Quasiperiodic Ising Models\hfil\thepage}
\makeatother
\title{\bf $\boldsymbol{Q}$-Dependent Susceptibilities in
\break Ferromagnetic Quasiperiodic $\boldsymbol{Z}$-Invariant
Ising Models \hfill}
\author{\bf Helen Au-Yang and Jacques~H.H.~Perk%
${}^{1,2}$\hfil}
\date{ }
\maketitle
\thispagestyle{plain}
\setcounter{footnote}{1}
\footnotetext{Department of Physics, Oklahoma State University,
Stillwater, OK 74078-3072, USA.}
\addtocounter{footnote}{1}
\footnotetext{Supported in part by NSF Grant No.\ PHY 01-00041.}
{\leftskip=5em\par\noindent\footnotesize
{\it Received September 17, 2004}
\par\noindent\hrulefill\vglue 5pt
\par\noindent
We study the $q$-dependent susceptibility $\chi({\bf q})$ of a series of
quasiperiodic Ising models on the square lattice. Several different kinds
of aperiodic sequences of couplings are studied, including the Fibonacci
and silver-mean sequences. Some identities and theorems are generalized
and simpler derivations are presented. We find that the $q$-dependent
susceptibilities are periodic, with the commensurate peaks of
$\chi({\bf q})$ located at the same positions as for the regular Ising
models. Hence, incommensurate everywhere-dense peaks can only occur in
cases with mixed ferromagnetic--antiferromagnetic interactions or if the
underlying lattice is aperiodic. For mixed-interaction models
the positions of the peaks depend strongly on the aperiodic
sequence chosen.
\par\noindent\hrulefill\vglue 5pt
\par\noindent{\sffamily\bfseries\footnotesize KEY WORDS:} Ising model;
$Z$-invariance; quasiperiodicity; golden ratio; silver mean; correlation
functions; wavevector-dependent susceptibility.
\par\leftskip=0pt}
\vglue 0pt

\advance\jot by 7pt
\section{Introduction}

In our most recent paper\cite{APpenta}, we have studied the $q$-dependent
susceptibility $\chi({\bf q})$ for a $Z$-invariant ferromagnetic Ising
model on Penrose tiles. (The $\chi({\bf q})$ is in many ways equivalent
to the structure function determining diffraction patterns.) We have
found that $\chi({\bf q})$ is aperiodic and has incommensurate peaks
which are everywhere dense, though only a limited number of them are
visible at temperatures far  away from the critical temperature. This is
very different from the behavior of $\chi({\bf q})$ in Fibonacci Ising
models defined on regular lattices\cite{AJPq, APmc1}, where
$\chi({\bf q})$ is periodic and has only commensurate peaks located at
the same positions as for the regular Ising models when the couplings
between the spins are ferromagnetic.

The periodicity of $\chi({\bf q})$, when the lattice is
regular, is due to the fact that we may write 
\begin{eqnarray} k_{\rm B}T\chi(q_x,q_y)=
\sum_{l,m} {\rm e}^{{\rm i}(q_x l+q_y m)} C(l,m)
\label{chi}\end{eqnarray}
where the average of the connected correlation function for two spins
with fixed separations $(l,m)$ is
\begin{eqnarray}
C(l,m)=\lim_{{\cal L}\to\infty}{\frac{1}{{\cal L}^{2}}}
\sum_{l',m'}\big[{\langle\sigma_{l',m'}\sigma_{l'+l,m'+m}\rangle}-
\langle\sigma_{l',m'}\rangle\langle\sigma_{l'+l,m'+m}\rangle\big].
\label{average}\end{eqnarray}
in which ${\cal L}$ denotes the number of rows and columns in the lattice,
so that ${\cal L}^2$ is the total number of spins. Since $l$ and $m$ are
integers, it is easily seen from (\ref{chi}) that the
$q$-dependent susceptibilities for such cases are periodic with periods
$2\pi$ in $q_x$ and $q_y$. When the lattice structure is quasi-periodic,
as in the case of the Penrose tiles studied in our previous
paper\cite{APpenta}, it is not possible to split the summation in the
susceptibility in this way and $\chi({\bf q})$ is no longer periodic.

In this paper, we want to examine the $q$-dependent susceptibility of some
other aperiodic ferromagnetic Ising models defined on regular
lattices, to find out if the Fibonacci Ising models are different from
other more general aperiodic models.

To be more specific, we consider the $Z$-invariant inhomogeneous Ising
model\cite{BaxZI,AP-ZI,Perkd, AJPq,APmc1} defined on a rectangular lattice
as shown in Fig.~1, and let either one of the sequences of rapidities,
$(u_n)_{n\in\mathbb Z}$ or $(v_m)_{m\in\mathbb Z}$ or both,
be certain aperiodic sequences. In doing so, we shall derive a number of
properties for these sequences which are part of the main results of this
paper.
\begin{figure}[tbp]
~\vskip0in\epsfclipon
\epsfxsize=0.85\hsize
\centerline{\epsfbox{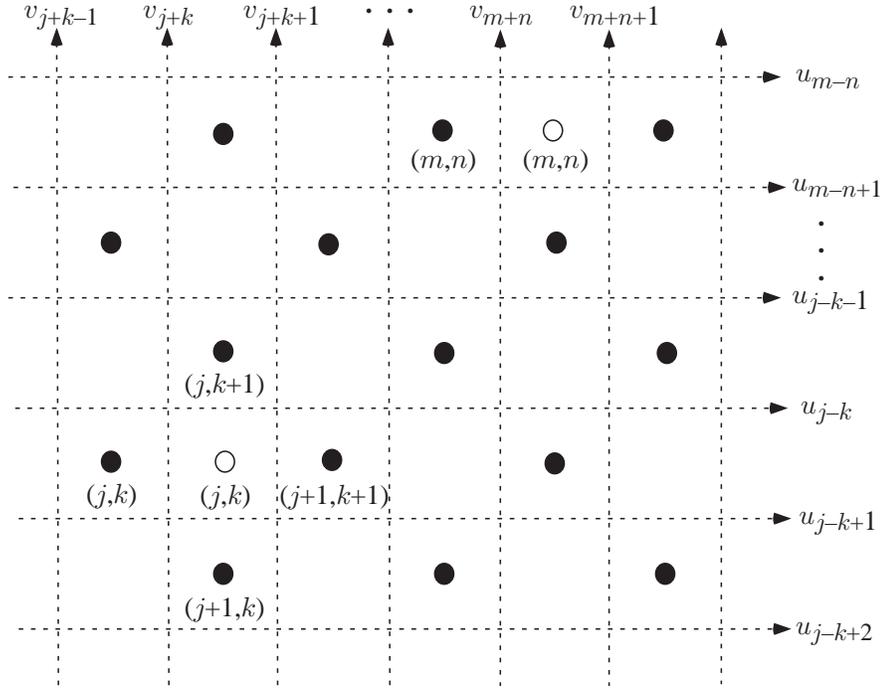}}
\vskip0.1in
\caption{Fig.~1. The lattice of a two-dimensional $Z$-invariant Ising
model: The rapidity lines on the medial graph are represented by oriented
dashed lines. The positions of the spins are indicated by small black
circles, the positions of two of the dual spins by white circles.}
\end{figure}
\par As before, the edge interactions are parametrized by (see Fig.~2)
\begin{eqnarray}
&&\sinh\big(2K(u_i,v_j)\big)=
k\,{\rm sc}(u_i-v_j,k')={\rm cs}\big(\lambda+v_j-u_i,k'\big),
\nonumber\\
&&\sinh\big(2{\bar K}(u_i,v_j)\big)
={\rm cs}(u_i-v_j,k')=k\,{\rm sc}\big(\lambda+v_j-u_i,k'\big),
\label{couplings}\nonumber\\
\end{eqnarray}
where $\lambda\equiv{\rm K}(k')$ is the complete elliptic integral of the
first kind, $k$ and $k'=\sqrt{1-k^2}$ are the elliptic moduli, which
are temperature variables, and they are the same for all couplings.
\begin{figure}[tbph]
~\vskip0in\epsfclipon
\epsfxsize=0.50\hsize
\centerline{\epsfbox{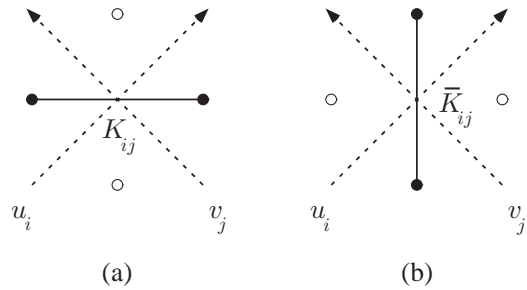}}
\vskip0.05in
\hskip0.1in\caption{Fig.~2. (a) Horizontal coupling $K_{ij}=K(u_i,v_j)$;
(b) Vertical coupling ${\bar K}_{ij}={\bar K}(u_i,v_j)$.}
\end{figure}
\section{Quasi-Periodic Sequences}
Quasi-periodic sequences were first used---within the related
context of the study of the specific heat of layered Ising models---by
Tracy \cite{Tr1,Tr2}. Even though the particular sequences used by
Tracy\cite{Tr2} may all be interesting, for some technical reasons we
shall consider here only the aperiodic sequences which were studied by de
Bruijn\cite{Bruijn0}. Let
\begin{eqnarray}\alpha_j\equiv{\textstyle \frac 1 2}
\big[(j+1)+\sqrt{(j+1)^2+4}\,\big],
\quad\hbox{for } j=0,1,2,\ldots,
\label{ratios}\end{eqnarray}
such that $\alpha_0=(1+\sqrt{5})/2$ is the golden ratio and
$\alpha_1=1+\sqrt{2}$ is the silver mean. Define for each $j$ a sequence
$(p_j(n))_{n\in\mathbb Z}$,
\begin{eqnarray}
p_j(n)\equiv\lfloor \gamma+(n+1)/\alpha_j\rfloor-\lfloor
\gamma+n/\alpha_j\rfloor,
\label{sequence}\end{eqnarray}
where $\lfloor x\rfloor$ is the largest integer $\le x$, and $\gamma$ is
a real number. In this paper, $\gamma$ is chosen such that
$\gamma+m/\alpha_j$ does not equal an integer for any $m$. Consequently,
the sequence in (\ref{sequence}) is not changed when the {\it floor}
($\lfloor x\rfloor$) in (\ref{sequence}) is replaced by {\it ceiling} or
{\it roof} ($\lceil x\rceil$: smallest integer $\ge x$). For the silver
mean sequence $(p_1(n))_{n\in\mathbb Z}$, we choose $\gamma\neq 
m+l\sqrt 2$ for all integers $m$ and $l$. More generally, it is
sufficient to require that $\gamma$ is not a solution of a quadratic
equation with integer coefficients.

It is shown by de Bruijn \cite{Bruijn0} that the
$(p_j(n))_{n\in\mathbb Z}$ are sequences of 0's and 1's, which may
also be easily shown by rewriting (\ref{sequence}) as 
\begin{equation}
p_j(n)=\lfloor x_n+1/\alpha_j\rfloor,\qquad
x_n=\{\gamma+n/\alpha_j\},\label{seq1}
\end{equation}
after decomposing $\gamma+n/\alpha_j$ into its integer and fractional
parts, i.e.,
\begin{eqnarray}
&\gamma+(n+1)/\alpha_j=\lfloor\gamma+n/\alpha_j\rfloor+
\{\gamma+n/\alpha_j\}+1/\alpha_j,\\
&\{x\}\equiv x-\lfloor x\rfloor,\qquad 0\le\{x\}<1.
\label{frac}\end{eqnarray} 
Since $\alpha_j>1$ and therefore $0\le x_n+1/\alpha_j<2$, it follows that
$p_j(n)=0$ if $0\le x_n+1/\alpha_j<1$ and
$p_j(n)=1$ if $1\le x_n+1/\alpha_j<2$. As $j$ increases (and so does
$\alpha_j$), the corresponding sequences $(p_j(n))_{n\in\mathbb Z}$
contain increasing numbers of zeros. In fact, for fixed $j$, the $p_j$'s
can be separated into blocks of $j+1$ digits (a one followed by $j$ zeros)
or $j+2$ digits (a one followed by $j+1$ zeros). Furthermore, it is also
shown by de Bruijn \cite{Bruijn0} that the production rule of replacing
each $1$ in a sequence $p_j$'s by a 1 followed by $j+1$ zeros and
replacing each 0 by a 1 followed by $j$ zeros produces a new sequence of
$p'_j$'s of the form (\ref{sequence}) with $\gamma\to\gamma'$ and
$\gamma'=-\{\gamma\}/\alpha_j$.

All these sequences are known to be aperiodic. Thus, if we let
$u_m=u_{\rm A}$ for $p_j(m)=1$, and $u_m=u_{\rm B}$ for $p_j(m)=0$, then
the sequence of line variables or rapidities $(u_m)_{m\in\mathbb Z}$ is
related to the sequence $(p_j(m))_{m\in\mathbb Z}$, and therefore
is also quasiperiodic. For $j=0$, the $p_0(m)$'s and the corresponding
$u_m$'s are Fibonacci sequences, and this case we have studied
earlier\cite{Tr1,AJPq,APmc1}. Likewise, we may also associate a
sequence of rapidities $(v_n)_{n\in\mathbb Z}$ to the sequence
$(p_j(n))_{n\in\mathbb Z}$. In this way, we can construct several
quasiperiodic $Z$-invariant Ising models on the square lattice. 

In order to calculate the average of the connected correlation functions,
$C(l,m)$ given by (\ref{average}), we need to generalize a result of Tracy
\cite{Tr1} for Fibonacci sequences. Tracy\cite{Tr2} mentions also some
other quasi-periodic sequences, for which---as far as we know---the
corresponding theorems are not yet available. But we can
generalize his result to general $j>0$ while simplifying his proof at
the same time.
\bigskip
\par\noindent{\bf Averages:} Following Tracy \cite{Tr1}, we let $N(n,m)$
be the number of 1's in the subsequence $p_j(m),\cdots,p_j(m+n-1)$ which
is also the number of $u_{\rm A}$'s in the subsequence
$u_m,\cdots,u_{m+n-1}$. 

Because the only allowed values of $p_j(n)$ are either 1 or 0, the
number of 1's among these $n$ consecutive terms of $p_j$'s is
\begin{eqnarray}
N(n,m)=\sum_{\ell=0}^{n-1}p_j(m+\ell)=
\lfloor \gamma+(m+n)/\alpha_j\rfloor-\lfloor
\gamma+m/\alpha_j\rfloor\cr
=\lfloor x_m+n/\alpha_j\rfloor=\lfloor x_m+\{n/\alpha_j\}\rfloor+\lfloor
n/\alpha_j\rfloor, \quad 
\label{freq}\end{eqnarray}
where $ x_m$ is defined in (\ref{seq1}) and $0\le x_m< 1$. Since
$0\le x_m+\{n/\alpha_j\}< 2$, we find
\begin{eqnarray}
N(n,m)=\cases{\begin{array}{llr}
\lfloor n/\alpha_j\rfloor &\hbox{for}&
x_m+\{n/\alpha_j\}< 1,\cr
\lfloor n/\alpha_j\rfloor+1 &\hbox{for}&
x_m+\{n/\alpha_j\}\ge 1.\end{array}} 
\label{case}\end{eqnarray}
Noting that $1/\alpha_j$ is irrational, we find from Kronecker's theorem
\cite{HW} that as $m$ varies from $-\infty$ to $\infty$, the $x_m$'s in
(\ref{seq1}) are distributed everywhere dense and uniformly between 0 and
1. Thus the probability of finding an $x_m$ with $x_m< 1-\{n/\alpha_j\}$
is $1-\{n/\alpha_j\}$, whereas the probability of
$x_m\ge 1-\{n/\alpha_j\}$ is $\{n/\alpha_j\}$. 
Consequently, we have proved the following theorem:

\medskip
\noindent{\bf Theorem 1:} An infinite quasiperiodic sequence
$(u_m)_{m\in\mathbb Z}$ defined by
$u_m=u_{\rm A}$ if $p_j(m)=1$ and $u_m=u_{\rm B}$ if $p_j(m)=0$ with the
$p_j$'s given by (\ref{sequence}) contains blocks of a single
$u_{\rm A}$ followed by either $j$ or $j+1$ $u_{\rm B}$'s. The number
of $u_{\rm A}$'s among $n$ consecutive $u$'s is either
$\lfloor n/\alpha_j\rfloor$ with probability $1-\{n/\alpha_j\}$ or
$\lfloor n/\alpha_j\rfloor+1$ with probability $\{n/\alpha_j\}$.

\medskip
\noindent We have thus generalized the result of Tracy \cite{Tr1} for
Fibonacci sequences (with $j=0$) to other cases ($j>0$), while also
simplifying the proof. 

\medskip
\noindent{\bf Sequences of Three Objects:}  Since each $p_j$ sequence is
quasiperiodic, if we shift a $p_j$ sequence by a certain number of digits
and subtract the shifted sequence from the original one, the resulting
sequence is also quasi-periodic, having three different values: $1$, $0$
and $-1$. Moreover, as a $p_j$ sequence consists of blocks of $j+1$ digits
with a one followed by $j$ zeros or blocks of $j+2$ digits with a one
followed by $j+1$ zeros, we find for $j\ne 0$ or
$\alpha_j\ne(1+\sqrt 5)/2$, that two consecutive terms in the original
$p_j$ sequence cannot be simultaneously 1. Consequently, if we let 
\begin{eqnarray}
q_j(\ell)=p_j(\ell+1)-p_j(\ell),\quad \ell\in\mathbb Z,
\label{sequence3}\end{eqnarray}
the average number of 1's (or $-1$'s) among $n$ consecutive numbers can be
easily evaluated. In this paper, we restrict ourselves to the sequences
(\ref{sequence3}) with $j\ge1$. Therefore, we work out the needed
probabilities next.

As it is indeed impossible to have both $p_j(\ell+1)$ and $p_j(\ell)$
equal to 1, we find $q_j(\ell)=1$ if $p_j(\ell+1)=1$; $q_j(\ell)=-1$ if
$p_j(\ell)=1$, and $0$ otherwise when $p_j(\ell+1)=p_j(\ell)=0$. Now we
let $u_m=u_{\rm A}$ if $q_j(m)=1$, $u_m=u_{\rm B}$ if $q_j(m)=0$, and
$u_m=u_{\rm C}$ if $q_j(m)=-1$. Consequently, the sequence of rapidities
$u_m$ is related to the sequence $q_j(m)$, and is therefore also
quasiperiodic.

Let the number $N_{\rm A}(n,m)$, ($N_{\rm B}(n,m)$ or $N_{\rm C}(n,m)$)
denote  the number of 1's, (0's or $-1$'s) in the subsequence
$q_j(m),\cdots$$,q_j(m\!+\!n\!-\!1)$, which is also the number of
$u_{\rm A}$, ($u_{\rm B}$ or $u_{\rm C}$) in the subsequence
$u_m,\cdots$$,u_{m+n-1}$. Since the number of 1's in
$q_j(m),\cdots$$,q_j(m\!+\!n\!-\!1)$ is equivalent to the number of 1's in
$p_j(m+1),\cdots$$,p_j(m+n)$, we find
\begin{eqnarray}
N_{\rm A}(n,m)&=&\lfloor
\gamma\!+\!(n\!+\!m\!+\!1)/\alpha_j\rfloor\!-\!\lfloor
\gamma\!+\!(m\!+\!1)/\alpha_j\rfloor\nonumber\\
&=&\lfloor x_{m+1}+n/\alpha_j\rfloor\nonumber\\
&=&\cases{\begin{array}{llr}
\lfloor n/\alpha_j\rfloor &\hbox{for}&
x_{m+1}< 1-\{n/\alpha_j\},\cr
\lfloor n/\alpha_j\rfloor+1 &\hbox{for}&
x_{m+1}\ge 1-\{n/\alpha_j\}.\end{array}} 
\label{freqa}\end{eqnarray}
cf.\ (\ref{freq}) and (\ref{case}).
Likewise, the number of $-1$'s in the new subsequence
$q_j(m),\cdots$$,q_j(m+n-1)$ is equivalent to the number of 1's in
the original $p_j(m),\cdots$$,p_j(m+n-1)$, and we find
\begin{eqnarray}
N_{\rm C}(n,m)&=&\lfloor
\gamma+(n+m)/\alpha_j\rfloor-\lfloor
\gamma+m/\alpha_j\rfloor\nonumber\\
&=&\lfloor x_{m}+n/\alpha_j\rfloor\nonumber\\
&=&\cases{\begin{array}{llr}
\lfloor n/\alpha_j\rfloor &\hbox{for}&
x_{m}< 1-\{n/\alpha_j\},\cr
\lfloor n/\alpha_j\rfloor+1 &\hbox{for}&
x_{m}\ge 1-\{n/\alpha_j\}.\end{array}} 
\label{freqc}\end{eqnarray}
Since the total must be $n$, we have
\begin{equation}
N_{\rm B}(n,m)=n-N_{\rm A}(n,m)-N_{\rm C}(n,m).
\end{equation}
Using (\ref{seq1}), we find
\begin{eqnarray}
x_{m+1}=\{ x_{m}+1/\alpha_j\}=\cases{\begin{array}{llr}
 x_{m}+1/\alpha_j &\hbox{for}&
x_m+1/\alpha_j< 1,\cr
 x_{m}+1/\alpha_j-1 &\hbox{for}&
x_m+1/\alpha_j\ge 1.\end{array}}
\label{freqd}\end{eqnarray}
\begin{figure}[tbp]
\vskip0in
\epsfxsize=1\hsize
\centerline{\epsfbox{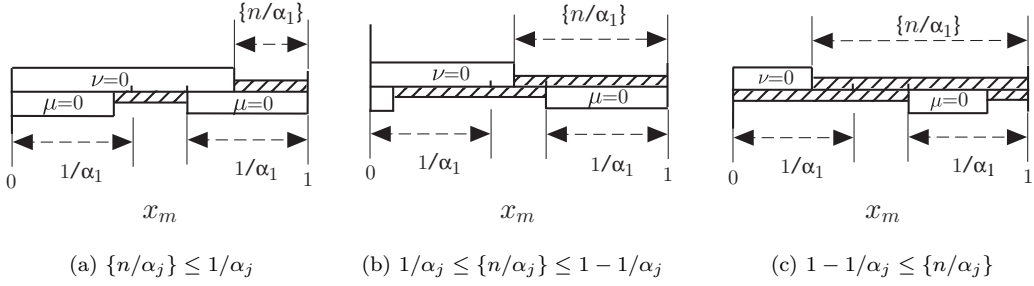}}
\hbox to\hsize{\hspace*{24pt}\scriptsize
(a) $\{n/\alpha_j\}\le 1/\alpha_j$\hfil
(b) $1/\alpha_j\le\{n/\alpha_j\}\le 1-1/\alpha_j$\hfil
(c) $1-1/\alpha_j\le\{n/\alpha_j\}$\hfil
\hspace*{-24pt}}
\vskip0.05in
\caption{Fig.~3. The regions of $x_m$ where
$N_{\rm A}(n,m)=\lfloor n/\alpha_j\rfloor+\mu$ and
$N_{\rm C}(n,m)=\lfloor n/\alpha_j\rfloor+\nu$ are shown for the
silver-mean case $j=1$. The segments where $\mu$ or $\nu=0$ are indicated
by thick white strips, while the segments where $\mu$ or $\nu=1$ are
indicated by narrow shaded strips. The $\mu$-strips are below and the
$\nu$-strips on top.}
\end{figure}
\par
In view of the above, let us write
\begin{eqnarray}
&&N_{\rm A}(n,m)=\lfloor n/\alpha_j\rfloor+\mu,\quad\hbox{with }
\mu=0\hbox{ or }1,\nonumber\\
&&N_{\rm C}(n,m)=\lfloor n/\alpha_j\rfloor+\nu,\quad\hbox{with }
\nu=0\hbox{ or }1.
\label{munu}\end{eqnarray}
Eqs.\ (\ref{freqa}), (\ref{freqc}), and (\ref{freqd}) determine the proper
choices of $\mu$ and $\nu$ as functions of $x_m$ and $\{n/\alpha_j\}$.
This is illustrated in Fig.~3 for the case $j=1$; the situation is
qualitatively the same for all $j\ge 1$.
We remind ourselves that the $x_m$ defined in (\ref{seq1}) is everywhere
dense and uniformly distributed in $[0,1)$, as $m$ runs from $-\infty$ to
$\infty$. Consequently, the choice $\nu=0$ is seen from (\ref{freqc}) and
(\ref{case}) to correspond to the segment in $[0,1]$ where the inequality
$0\le x_{m}< 1-\{n/\alpha_j\}$ is satisfied, while $\nu=1$ is given by its
complement satisfying $1-\{n/\alpha_j\}\le x_{m}< 1$. Using
(\ref{freqd}), we find $0\le x_{m+1}<1-\{n/\alpha_j\}$ is equivalent to
both $1-1/\alpha_j\le x_{m}< 2-\{n/\alpha_j\}-1/\alpha_j$ and
$0\le x_{m}< 1-\{n/\alpha_j\}-1/\alpha_j$. Since $0\le x_m<1$, the
second inequality cannot be satisfied if $1-\{n/\alpha_j\}-1/\alpha_j<0$,
which is the defining condition for Fig.~3$\,$(c); here $\mu=0$ is the
segment where $1-1/\alpha_j\le x_{m}< 2-\{n/\alpha_j\}-1/\alpha_j$; its
complement $\mu=1$, however, consists of two disjunct segments.
Cases with $1-\{n/\alpha_j\}-1/\alpha_j> 0$ are shown in Fig.~3$\,$(a) and
(b); here $\mu=0$ consists of two disjunct segments satisfying
$1-1/\alpha_j\le x_{m}< 1$ or $0\le x_{m}< 1-\{n/\alpha_j\}-1/\alpha_j$,
while its complement $\mu=1$ is now just one segment given by
$1-\{n/\alpha_j\}-1/\alpha_j\le x_{m}< 1-1/\alpha_j$.

Let $P(\mu',\nu')$, for $\mu',\nu'=0,1$, denote the joint probability for
having both $N_{\rm A}(n,m)=\lfloor n/\alpha_j\rfloor+\mu'$ and
$N_{\rm C}(n,m)=\lfloor n/\alpha_j\rfloor+\nu'$. Then $P(\mu',\nu')$ is
the total length of the intersection of the segment or segments where
$\mu=\mu'$ with the segment where $\nu=\nu'$. The results are different
for the three different regions of $\{n/\alpha_j\}$. We find
\begin{eqnarray}
\hspace{-30pt}&&\left.\begin{array}{l}
P(0,0)=1-2\{n/\alpha_j\}\cr
P(1,0)=P(0,1)=\{n/\alpha_j\}\cr
P(1,1)=0
\end{array}\!\right\}
\hbox{ if }\{n/\alpha_j\}\le 1/\alpha_j,
\label{prob1}\\
\hspace{-30pt}&&\left.\begin{array}{l}
P(0,0)=1-\{n/\alpha_j\}-1/\alpha_j\cr
P(1,0)=P(0,1)=1/\alpha_j\cr
P(1,1)=\{n/\alpha_j\}-1/\alpha_j
\end{array}\!\right\}
\hbox{ if }1/\alpha_j\le\{n/\alpha_j\}\le 1-1/\alpha_j,
\label{prob2}\\
\hspace{-30pt}&&\left.\begin{array}{l}
P(0,0)=0\cr
P(1,0)=P(0,1)=1-\{n/\alpha_j\}\cr
P(1,1)=2\{n/\alpha_j\}-1
\end{array}\!\right\}
\hbox{ if }\{n/\alpha_j\}\ge 1-1/\alpha_j.
\label{prob3}\end{eqnarray}

\noindent{\bf Remark:} Both Theorem 1 for the two-object case and
Eqs.\ (\ref{munu}) through (\ref{prob3}) for the three-object case have
a reflection symmetry under the formal replacement $n\to-n$, ($n>0$).
Since $\alpha_j$ is irrational, this means we have to replace
$\{n/\alpha_j\}\to\{-n/\alpha_j\}=1-\{n/\alpha_j\}$,
so that $P(\mu,\nu)\to P(1-\mu,1-\nu)$. Also,
$-\lfloor -n/\alpha_j\rfloor=\lceil n/\alpha_j\rceil=
\lfloor n/\alpha_j\rfloor+1$, and (\ref{munu}) is
to be replaced by $N_{\rm A}(-n,m)=\lfloor n/\alpha_j\rfloor+1-\mu$,
$N_{\rm C}(-n,m)=\lfloor n/\alpha_j\rfloor+1-\nu$. Therefore, we indeed
have that the probability distribution is invariant under reflections.
It is also translationally invariant, as (\ref{munu})--(\ref{prob3}) are
independent of $m$.
\section{Correlations}
The spin-spin correlation function in the inhomogeneous $Z$-invariant
Ising model has been shown by Baxter \cite {BaxZI} to depend only on the
elliptic modulus $k$ and the rapidity variables, $u$'s and $v$'s, of
rapidity lines that are sandwiched between the two spins.  Particularly,
for $-l\le m\le l$, when the arrows of all these relevant rapidity lines
are pointing to the same side of the line joining the two spins (see
Fig.~1), we have---according to the rule in ref.\ \citen{AP-ZI}---the
result
\begin{eqnarray}
\langle\sigma_{j,k}\sigma_{j+l,k+m}\rangle=g^{\vphantom{*}}_{2l}(
u_{j-k+1},\ldots,u_{l-m+j-k},v_{j+k},\ldots,v_{l+m+j+k-1}),
\label{corz1}\\
\langle\mu_{j,k}\mu_{j+l,k+m}\rangle=g^*_{2l}(
u_{j-k+1},\ldots,u_{l-m+j-k},v_{j+k+1},\ldots,v_{l+m+j+k}),
\label{dualcorz1}
\end{eqnarray}
while for $-m\le l\le m$, when the arrows of the vertical rapidity
lines and the arrows of the horizontal rapidity lines are pointing to
opposite sides of the joining line, we find
\begin{eqnarray}
\hspace{-30pt}&&\langle\sigma_{j,k}\sigma_{j+l,k+m}\rangle=\nonumber\\
\hspace{-30pt}&&\qquad g^{\vphantom{*}}_{2m}(
u_{j-k+l-m+1},\ldots,u_{j-k},\lambda+ v_{j+k},
\ldots,\lambda+v_{l+m+j+k-1}),
\label{corz2}\\
\hspace{-30pt}&&\langle\mu_{j,k}\mu_{j+l,k+m}\rangle=\nonumber\\
\hspace{-30pt}&&\qquad g^*_{2m}(
u_{j-k+l-m+1},\ldots,u_{j-k},\lambda+v_{j+k+1}.
\ldots,\lambda+v_{l+m+j+k}), 
\label{dualcorz2}\end{eqnarray}
Here $\lambda\equiv{\rm K}(k')$ is a complete elliptic integral of the
first kind. Note that the explicit dependence of $\lambda$, $g$ and
$g^*$ on the elliptic modulus $k$ is dropped, but it should still be
understood to be implicitly present. Also, the $\mu\equiv\sigma^*$ stand
for dual spins on the dual lattice, which is at the dual temperature.

As pointed out first by Baxter, \cite {BaxZI} the universal functions
$g_{2l}^{\vphantom{*}}$ and $g_{2l}^*$ have ``permutation symmetry"
(meaning they are invariant under all permutations of the rapidities) and
the ``difference property" (which implies a translation invariance when
shifting all the rapidities by the same amount $v^{(0)}$). The functions
$g_{2l}^{\vphantom{*}}$ and $g_{2l}^*$ for $l>1$ can be obtained
iteratively \cite{AJPq,APmc1,AP-ZI}. The final technical point is to
explain how the averaging in (\ref{average}) is done.

\subsection{Averaging}

In this paper, we shall consider quasiperiodic sequences which are
either sequences of two objects:
\begin{eqnarray}
u_{m}=\cases{\begin{array}{llr}
u_{\rm A} &\hbox{if}&p_j(m)=1,\cr
u_{\rm B} &\hbox{if}&p_j(m)=0,\end{array}}\quad
v_{m}=\cases{\begin{array}{llr}
v_{\rm A} &\hbox{if}&p_j(m)=1,\cr
v_{\rm B} &\hbox{if}&p_j(m)=0,\end{array}}
\label{abseq}\end{eqnarray}
for fixed $j\ge 0$, or
sequences of three objects:
\begin{eqnarray}
u_{m}=\cases{\begin{array}{lll}
u_{\rm A} &\hbox{if}&q_j(m)=1,\cr
u_{\rm B} &\hbox{if}&q_j(m)=0,\cr
u_{\rm C} &\hbox{if}&q_j(m)=-1,\end{array}}\quad v_m=v, \quad j\ge 1. 
\label{abcseq}\end{eqnarray}

To evaluate $C(l,m)$ and $C^*(l,m)$ for $|m|\le l$, we use (\ref{corz1})
and (\ref{dualcorz1}). It is easily seen from these equations that there
are $l-m$ horizontal rapidity lines $u$ and $l+m$ vertical lines $v$
sandwiched between the two spins.

For the two-object sequences in (\ref{abseq}), we find from Theorem 1
that the number of $u_{\rm A}$'s among the $l-m$ consecutive $u$'s is
either $\lfloor\cn\rfloor$ with probability $1-\{\cn\}$ or
$\lfloor\cn\rfloor+1$ with probability $\{\cn\}$ where
$\cn=(l-m)/\alpha_j$, while the number of $v_{\rm A}$'s among the $l+m$
consecutive $v$'s is either $\lfloor\cm\rfloor$ with probability
$1-\{\cm\}$ or $\lfloor\cm\rfloor+1$ with probability $\{\cm\}$ in which
$\cm=(l+m)/\alpha_j$. Consequently, the averaged connected
correlation function in (\ref{average}) for $|m|\le l$ becomes
\begin{eqnarray}
C(l,m)&=&(1-\{\cn\})(1-\{\cm\})\,
\bg[\lfloor\cn\rfloor,l-m-\lfloor\cn\rfloor,
\lfloor\cm\rfloor,l+m-\lfloor\cm\rfloor]\nonumber\\
&+&(1-\{\cn\})\{\cm\}\,\bg[\lfloor\cn\rfloor, l-m-\lfloor \cn\rfloor,
\lfloor\cm\rfloor+1,l+m-\lfloor\cm\rfloor-1]\nonumber\\
&+&\{\cn\}(1-\{\cm\})\,\bg[\lfloor\cn\rfloor+1, l-m-\lfloor \cn\rfloor-1,
\lfloor\cm\rfloor,l+m-\lfloor\cm\rfloor]\nonumber\\
&+&\{\cn\}\{\cm\}\,\bg[\lfloor\cn\rfloor+1, l-m-\lfloor \cn\rfloor-1,
\lfloor\cm\rfloor+1,l+m-\lfloor\cm\rfloor-1]\nonumber\\
&-&\langle\sigma\rangle^2,
\label{avcor1}\end{eqnarray}
where
\begin{eqnarray}
&&\cn\equiv(l-m)/\alpha_j,\qquad \cm\equiv(l+m)/\alpha_j,\\
&&\bg[m_3,m_2,m_1,m_0]\equiv\nonumber\\
&&\qquad\qquad g(\overbrace{u_{\rm A},\dots,u_{\rm A}}^{m_3},
\overbrace{u_{\rm B},\dots,u_{\rm B}}^{m_2},
\overbrace{v_{\rm A},\dots,v_{\rm A}}^{m_1},
\overbrace{v_{\rm B},\dots,v_{\rm B}}^{m_0})
\label{bg}\end{eqnarray}
and $\langle\sigma\rangle=0$, as $T\ge T_{\rm c}$.
The averaged correlation $C^*(l,m)$ of the disorder variables, which is
also the correlation function for $T\le T_{\rm c}$, can be obtained from
the above equations simply by replacing $g$ by $g^*$ and
$\langle\sigma\rangle$ by $(1-k^{-2})^{1/8}$.

For the three-object sequences in (\ref{abcseq}), the numbers of
$u_{\rm A}$'s and $u_{\rm C}$'s among the $l-m$ consecutive $u$'s are
given by (\ref{freqa}) and (\ref{freqc}), and the averaged connected
correlation in (\ref{average}) can be evaluated using (\ref{prob1})
through (\ref{prob3}). We find, for $\{\cn\}\le 1/\alpha_j$,
\begin{eqnarray}
C(l,m)&=&
(1-2\{\cn\})\,\hg[\lfloor\cn\rfloor,l-m-2\lfloor\cn\rfloor,
\lfloor\cn\rfloor,l+m]\nonumber\\
&+&\{\cn\}\,\hg[\lfloor\cn\rfloor, l-m-2\lfloor \cn\rfloor-1,
\lfloor\cn\rfloor+1,l+m]\nonumber\\
&+&\{\cn\}\,\hg[\lfloor\cn\rfloor+1, l-m-2\lfloor \cn\rfloor-1,
\lfloor\cn\rfloor,l+m]\nonumber\\
&-&\langle\sigma\rangle^2,
\label{avcor3a}\end{eqnarray}
where
\begin{eqnarray}
&&\hg[m_3,m_2,m_1,m_0]\equiv\nonumber\\
&&\qquad\qquad g(\overbrace{u_{\rm A},\dots,u_{\rm A}}^{m_3},
\overbrace{u_{\rm B},\dots,u_{\rm B}}^{m_2},
\overbrace{u_{\rm C},\dots,u_{\rm C}}^{m_1},
\overbrace{v,\dots,v}^{m_0}).
\label{hg}\end{eqnarray}
For $1/\alpha_j\le \{\cn\}\le 1-1/\alpha_j$, we find
\begin{eqnarray}
C(l,m)&=&
(1-\{\cn\}-1/\alpha_j)\,\hg[\lfloor\cn\rfloor,
l-m-2\lfloor \cn\rfloor,
\lfloor\cn\rfloor,l+m]\nonumber\\
&+&(1/\alpha_j)\,\hg[\lfloor\cn\rfloor, l-m-2\lfloor \cn\rfloor-1,
\lfloor\cn\rfloor+1,l+m]\nonumber\\
&+&(1/\alpha_j)\,\hg[\lfloor\cn\rfloor+1, l-m-2\lfloor \cn\rfloor-1,
\lfloor\cn\rfloor,l+m]\nonumber\\
&+&(\{\cn\}-1/\alpha_j)\,\hg[\lfloor\cn\rfloor+1, l-m-2\lfloor
\cn\rfloor-2,
\lfloor\cn\rfloor+1,l+m]\nonumber\\
&-&\langle\sigma\rangle^2,
\label{avcor3b}\end{eqnarray}
whereas, for $\{\cn\}\ge 1-1/\alpha_j $, 
\begin{eqnarray}
C(l,m)&=&
(1-\{\cn\})\,\hg[\lfloor\cn\rfloor,l-m-2\lfloor
\cn\rfloor-1,\lfloor\cn\rfloor+1,l+m]\nonumber\\
&+&(1-\{\cn\})\,\hg[\lfloor\cn\rfloor+1,l-m-2\lfloor \cn\rfloor-1,
\lfloor\cn\rfloor,l+m]\nonumber\\
&+&(2\{\cn\}-1)\,\hg[\lfloor\cn\rfloor+1,l-m-2\lfloor \cn\rfloor-2,
\lfloor\cn\rfloor+1,l+m]\nonumber\\
&-&\langle\sigma\rangle^2.
\label{avcor3c}\end{eqnarray}
Again, the formulae for $C^*(l,m)$ are similar, cf.\ the
discussion below (\ref{bg}). Also, it is easily verified that we have the
general inversion symmetry
\begin{eqnarray}
C(-l,-m)=C(l,m),\qquad C^*(-l,-m)=C^*(l,m),
\label{inversion}\end{eqnarray}
valid for all values of $l$ and $m$. Hence, we have now the results for
$|m|\le |l|$.

To evaluate $C(-m,l)$ and $C^*(-m,l)$ for $|m|\le l$, we let $l\to-m$ and
$m\to l$ in (\ref{corz2}) and (\ref{dualcorz2}), and find that there are
$l+m$ horizontal lines $u$ and $l-m$ vertical lines $v$ sandwiched
between the two spins. 

If $u_m$ and $v_n$ are given by (\ref{abseq}), Theorem 1 can again be
used to find the average number of $u_{\rm A}$'s among the $l+m$
consecutive $u$'s and the average number of $v_{\rm A}+\lambda$'s among
the $l-m$ consecutive $v$'s. As a consequence the averaged connected
correlation function in (\ref{average}) for $|m|\le l$ becomes
\begin{eqnarray}
C(-m,l)&=&
(1-\{\cm\})(1-\{\cn\})\,g'[\lfloor\cm\rfloor,l+m-\lfloor\cm\rfloor,
\lfloor\cn\rfloor, l-m-\lfloor\cn\rfloor]\nonumber\\
&+&\{\cm\}(1-\{\cn\})\,g'[\lfloor\cm\rfloor+1,l+m-\lfloor\cm\rfloor-1,
\lfloor\cn\rfloor,l-m-\lfloor \cn\rfloor]\nonumber\\
&+&(1-\{\cm\})\{\cn\}\,g'[\lfloor\cm\rfloor,l+m-\lfloor\cm\rfloor,
\lfloor\cn\rfloor+1, l-m-\lfloor \cn\rfloor-1]\nonumber\\
&+&\{\cm\}\{\cn\}\,g'[\lfloor\cm\rfloor+1,l+m-\lfloor\cm\rfloor-1,
\lfloor\cn\rfloor+1, l-m-\lfloor \cn\rfloor-1]\nonumber\\
&-&\langle\sigma\rangle^2,
\label{avcor2}\end{eqnarray}
where
\begin{eqnarray}
&&g'[m_3,m_2,m_1,m_0]\equiv\nonumber\\
&&\qquad
g(\overbrace{\vphantom{\lambda}u_{\rm A},\dots,u_{\rm A}}^{m_3},
\overbrace{\vphantom{\lambda}u_{\rm B},\dots,u_{\rm B}}^{m_2},
\overbrace{\lambda+v_{\rm A},\dots,\lambda+v_{\rm A}}^{m_1},
\overbrace{\lambda+v_{\rm B},\dots,\lambda+v_{\rm B}}^{m_0}).\nonumber\\
\label{gp}\end{eqnarray}
For the three-object sequences in (\ref{abcseq}), the average numbers of
$u_{\rm A}$'s and $u_{\rm C}$'s among the $l+m$ consecutive $u$'s are
given by (\ref{prob1}) through (\ref{prob3}).
For $|m|\le l$ and $\{\cm\}\le 1/\alpha_j$, we find
\begin{eqnarray}
C(-m,l)&=&
(1-2\{\cm\})\,\hg'[\lfloor\cm\rfloor, l+m-2\lfloor\cm\rfloor,
\lfloor\cm\rfloor,l-m]\nonumber\\
&+&\{\cm\}\,\hg'[\lfloor\cm\rfloor, l+m-2\lfloor \cm\rfloor-1,
\lfloor\cm\rfloor+1,l-m]\nonumber\\
&+&\{\cm\}\,g[\lfloor\cm\rfloor+1, l+m-2\lfloor \cm\rfloor-1,
\lfloor\cm\rfloor,l-m]\nonumber\\
&-&\langle\sigma\rangle^2,
\label{avcor3aa}\end{eqnarray}
while, for $1/\alpha_j\le \{\cm\}\le 1-1/\alpha_j$,
\begin{eqnarray}
C(-m,l)&=&
(1-\{\cm\}-1/\alpha_j)\,
\hg'[\lfloor\cm\rfloor, l+m-2\lfloor
\cm\rfloor,\lfloor\cm\rfloor,l-m]\nonumber\\
&+&(1/\alpha_j)\,\hg'[\lfloor\cm\rfloor, l+m-2\lfloor \cn\rfloor-1,
\lfloor\cm\rfloor+1,l-m]\nonumber\\
&+&(1/\alpha_j)\,\hg'[\lfloor\cm\rfloor+1, l+m-2\lfloor \cm\rfloor-1,
\lfloor\cm\rfloor,l-m]\nonumber\\
&+&(\{\cm\}-1/\alpha_j)\,\hg'[\lfloor\cm\rfloor+1, l+m-2\lfloor
\cm\rfloor-2,
\lfloor\cm\rfloor+1,l-m]\nonumber\\
&-&\langle\sigma\rangle^2,
\label{avcor3bb}\end{eqnarray}
whereas, for $\{\cm\}\ge 1-1/\alpha_j $, 
\begin{eqnarray}
C(-m,l)&=&
(1-\{\cm\})\,\hg'[\lfloor\cm\rfloor,
l+m-2\lfloor
\cm\rfloor-1,\lfloor\cm\rfloor+1,l-m]\nonumber\\
&+&(1-\{\cm\})\,\hg'[\lfloor\cm\rfloor+1, l+m-2\lfloor \cm\rfloor-1,
\lfloor\cm\rfloor,l-m]\nonumber\\
&+&(2\{\cm\}-1)\,\hg'\lfloor\cm\rfloor+1, l+m-2\lfloor \cm\rfloor-2,
\lfloor\cm\rfloor+1,l-m]\nonumber\\
&-&\langle\sigma\rangle^2,
\label{avcor3cc}\end{eqnarray}
where
\begin{eqnarray}
&&\hg'[m_3,m_2,m_1,m_0]\equiv\nonumber\\
&&\qquad g(\overbrace{\vphantom{\lambda}u_{\rm A},\dots,u_{\rm A}}^{m_3},
\overbrace{\vphantom{\lambda}u_{\rm B},\dots,u_{\rm B}}^{m_2},
\overbrace{\vphantom{\lambda}u_{\rm C},\dots,u_{\rm C}}^{m_1},
\overbrace{\lambda+v,\dots,\lambda+v}^{m_0}).
\label{hgp}\end{eqnarray}

In view of (\ref{inversion}) and the discussion below (\ref{bg}),
we have now obtained a complete set of formulae for $C(l,m)$ and
$C^*(l,m)$. We can use difference equations to obtain, by iteration, all
needed $g$'s and $g^*$'s. The details of such calculations are in our
previous work\cite{APmc1,APmc2}, and will not be presented here. Since
the various $g[m_3,m_2,m_1,m_0]$'s are obtained iteratively from $g$'s and
$g^*$'s with smaller $m_i$'s, it is necessary to evaluate the
$g[m_3,m_2,m_1,m_0]$'s for almost all $m_i$ such that $0\le m_i\le N$ even
though for each fixed $\alpha_j$, only a fraction of these $g$'s
are needed. In spite of powerful modern computers, these calculations are
still quite time consuming. Therefore, it is more economical to obtain the
correlations for all different $j$'s studied in one shot.

In the next section we shall be a little more specific.
If the four above  rapidities $(u_{\rm A},u_{\rm B},v_{\rm A},v_{\rm B})$
or $(u_{\rm A},u_{\rm B},u_{\rm C},v)$ are chosen to be multiples of
$\lambda/4$ in a certain way (details will be given later), we can use the
permutation property to arrange the rapidities in $g$ and $g^*$ in
descending order, and then use the difference property to make the
smallest rapidity identically equal to zero. Then, all functions $g$ and
$g^*$ in (\ref{bg}), (\ref{gp}), (\ref{hg}) and (\ref{hgp}) can be
brought to the form
\begin{eqnarray}
g[m_3,m_2,m_1,m_0]=
g(\overbrace{{{\textstyle \frac 3 4}\lambda},
\dots,{{\textstyle \frac 3 4}\lambda}}^{m_3},
\overbrace{{{\textstyle \frac 1 2}\lambda},
\dots,{{\textstyle \frac 1 2}\lambda}}^{m_2},
\overbrace{{{\textstyle \frac 1 4}\lambda},
\dots,{{\textstyle \frac 1 4}\lambda}}^{m_1},
\overbrace{\vphantom{{\textstyle \frac 1 4}}0,\dots,0}^{m_0}),
\label{g}\end{eqnarray}
possibly permuting the $m_i$'s.

\section{Wavevector-Dependent Susceptibility}
Since the correlation functions decay exponentially ($T\ne T_{\rm c}$), we
need to put all the terms that have approximately the same order of
magnitude together. More specifically we write
\begin{eqnarray}
\bar\chi(q_x,q_y)\equiv k_{\rm B}T\chi(q_x,q_y)=C(0,0)
+2\sum_{l=1}^{\infty}\, S_l,\quad(\hbox{with }C(0,0)=1),\cr
S_l=\sum_{m=1-l}^l \left[C(l,m)\,\cos(q_x l+q_y m)
+C(-m,l)\,\cos(-q_x m+q_y l)\right], 
\label{chis}\end{eqnarray}
where $S_l$ contains the  correlations of the top and right edges of the
square whose four corners are $(\pm l,\pm l)$. The above cosines result
from the use of the inversion symmetry (\ref{inversion}) in order to
include the contributions of the other two edges. For $T$ away from
$T_{\rm c}$, only a few $S_l$ for $l$ small are numerically significant.
As $T\to T_{\rm c}$, more and more terms need be included. This way the
$q$-dependent susceptibility can now be evaluated for different cases.

\subsection{Sequences of Two Objects, Example I} 
We shall first consider some quasiperiodic sequences of two objects. 
Let the sequence of rapidity lines be defined by (\ref{abseq}) with the
particular values
\begin{eqnarray}
u_{\rm A}=3\lambda/4,\quad u_{\rm B}=2\lambda/4,
\quad v_{\rm A}=\lambda/4,\quad v_{\rm B}=0.
\label{rapid2}\end{eqnarray}
Comparing (\ref{bg}) and (\ref{gp}) with (\ref{g}), we find 
\begin{eqnarray}
\bg[m_3,m_2,m_1,m_0]=g[m_3,m_2,m_1,m_0],\nonumber\\
\quad g'[m_1,m_0,m_3,m_2]=g[m_3,m_2,m_1,m_0],
\label{bgg}\end{eqnarray}
where the permutation property and the difference property \cite{BaxZI}
are used for the second identity. Now, comparing (\ref{avcor1}) with
(\ref{avcor2}), we find 
\begin{eqnarray}
C(-m,l)=C(l,m),\quad C^*(-m,l)=C^*(l,m).
\label{4fold}\end{eqnarray}
From (\ref{4fold}), we find that the $q$-dependent susceptibility must
have fourfold rotational symmetry.

We can calculate the $q$-dependent susceptibility for fixed
$T\ne T_{\rm c}$, ($k\ne1$), to arbitrary precision using an algorithm of
polynomial complexity. For this purpose, we use quadratic difference
equations \cite{AJPq,APmc1,AP-ZI,Perkd,APmc2} to numerically evaluate the
averaged correlation functions given by (\ref{avcor1}), (\ref{bgg}) and
(\ref{4fold}) for $T>T_{\rm c}$, and replace $g$ and
$\langle\sigma\rangle\equiv0$ by $g^*$ and
$\langle\sigma\rangle=(1-k^{-2})^{1/8}$ for $T<T_{\rm c}$. We
have used Maple software for this, as higher and higher precision
arithmetic is needed closer and closer to $T_{\rm c}$. Substituting the
results into (\ref{chis}), we obtain the $q$-dependent susceptibility at
different temperatures. We shall present our results mostly in density
plots to get an overview of the full $(q_x,q_y)$-dependence. Our results,
however, are far more accurate than these plots suggest.

In Fig.~4, we show four density plots of $1/\chi({\bf q})$ for $j=0$,
1, 2 or 3 and $-2\pi<q_x,q_y<2\pi$, at the one temperature $T>T_{\rm c}$
for which the above-$T_{\rm c}$ correlation length
$\xi\approx8$.\footnote{More precisely, $\xi$ is the row correlation
length of the uniform and symmetric square-lattice Ising model with the
same value of modulus $k=\big(\cosh^2(\frac{1}{2}\xi^{-1})\pm
[\cosh^4(\frac{1}{2}\xi^{-1})-1]^{1/2}\big)^2$, with minus for
$T>T_{\rm c}$. For $T<T_{\rm c}$, we must choose plus, while
$\xi$ is then twice the actual row correlation length.\cite{MW}}
(In the density plots, darker means a relatively larger value of
$\chi({\bf q})$, and $\mathsf x\equiv q_x$, $\mathsf y\equiv q_y$.) We
find that there is no incommensurate behavior, for all different values
of $\alpha_j$ with $j\ge 0$ and at arbitrary temperature. The peaks of
$\chi({\bf q})$ are at the commensurate positions of the ordinary Ising
model, i.e.\ $(q_x,q_y)=(2\pi m,2\pi n)$ where $m$ and $n$ are any
integers. We also find that $\chi({\bf q})$ is indeed invariant under
$90^{\circ}$ rotation.
\begin{figure}[htbp]
\vskip0.3in\hskip-2pt\epsfclipon
\epsfxsize=0.475\hsize\epsfbox{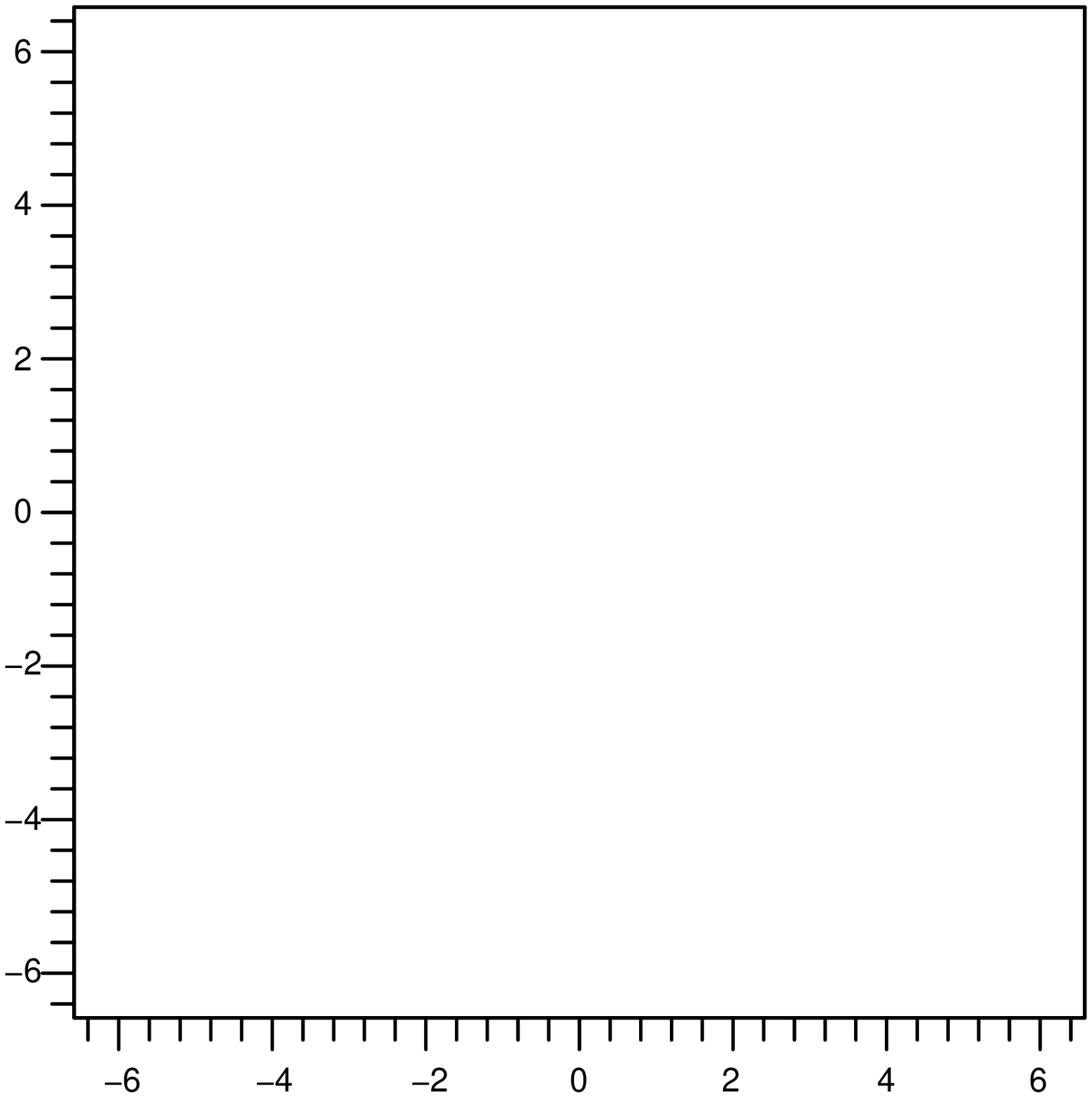}\hspace*{-0.4225\hsize}%
\epsfxsize=0.370\hsize\raisebox{0.0517\hsize}{\epsfbox{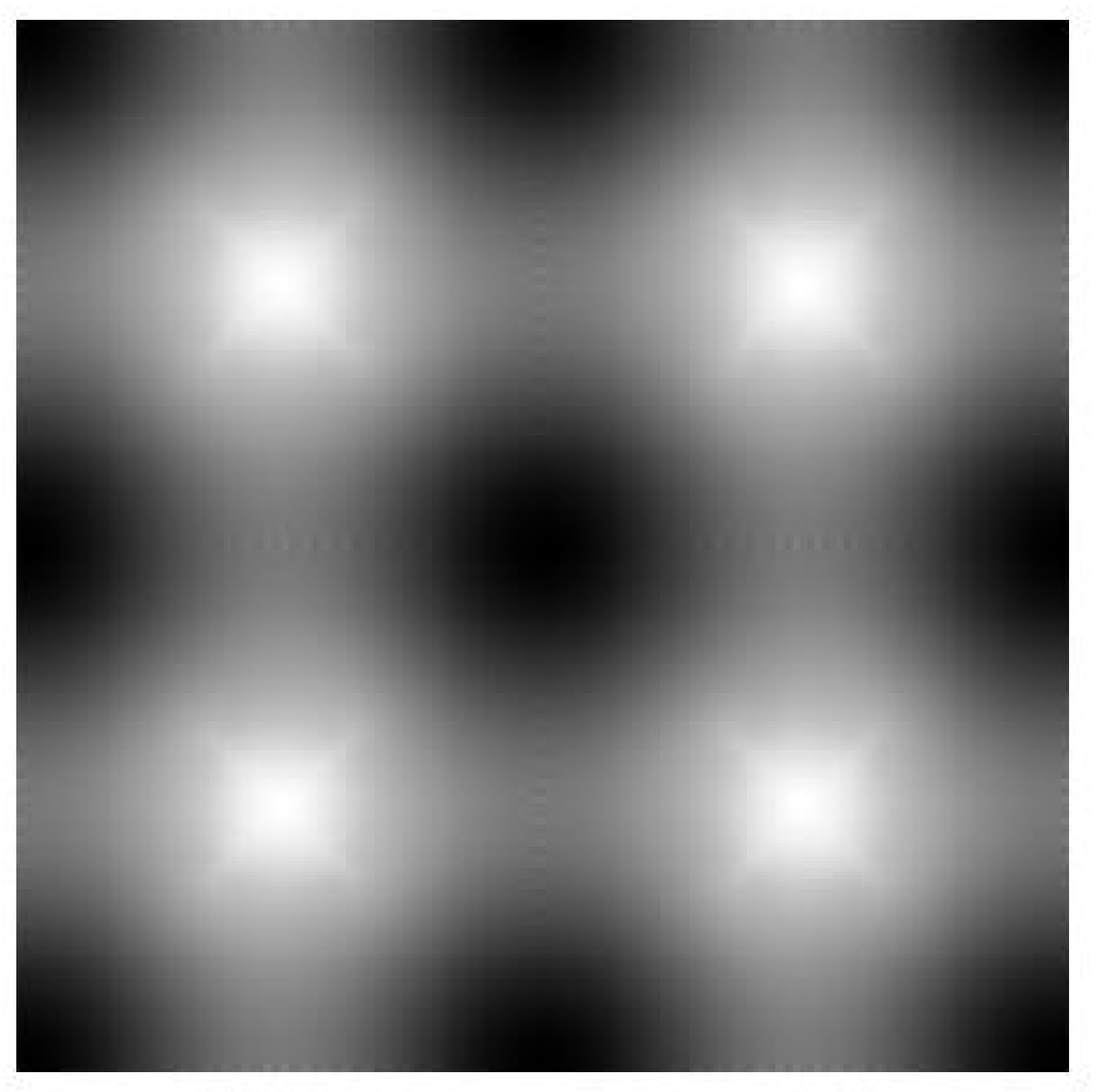}}\hfil
\epsfxsize=0.475\hsize\epsfbox{fig4678z.eps}\hspace*{-0.4225\hsize}%
\epsfxsize=0.370\hsize\raisebox{0.0517\hsize}{\epsfbox{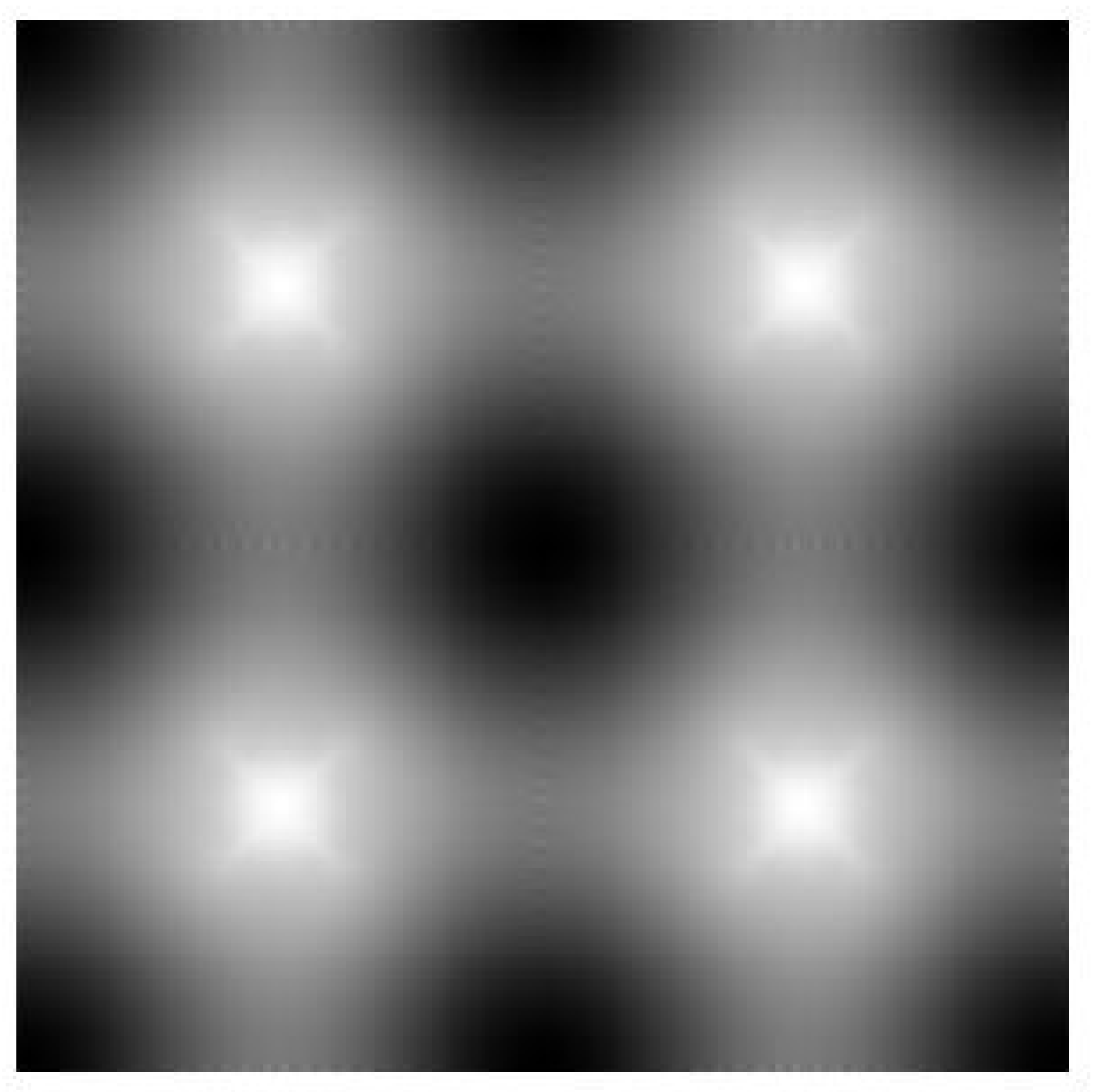}}
\hbox to\hsize{\hspace*{12pt}\scriptsize
(a) golden ratio: $j=0$ and $\alpha_0=(\sqrt 5+1)/2$\hfil\hspace*{-8pt}
(b) silver mean: $j=1$ and $\alpha_1=\sqrt 2+1$\hfil
\hspace*{-48pt}}
\vskip0.2in\hskip0pt
\epsfxsize=0.475\hsize\epsfbox{fig4678z.eps}\hspace*{-0.4225\hsize}%
\epsfxsize=0.370\hsize\raisebox{0.0517\hsize}{\epsfbox{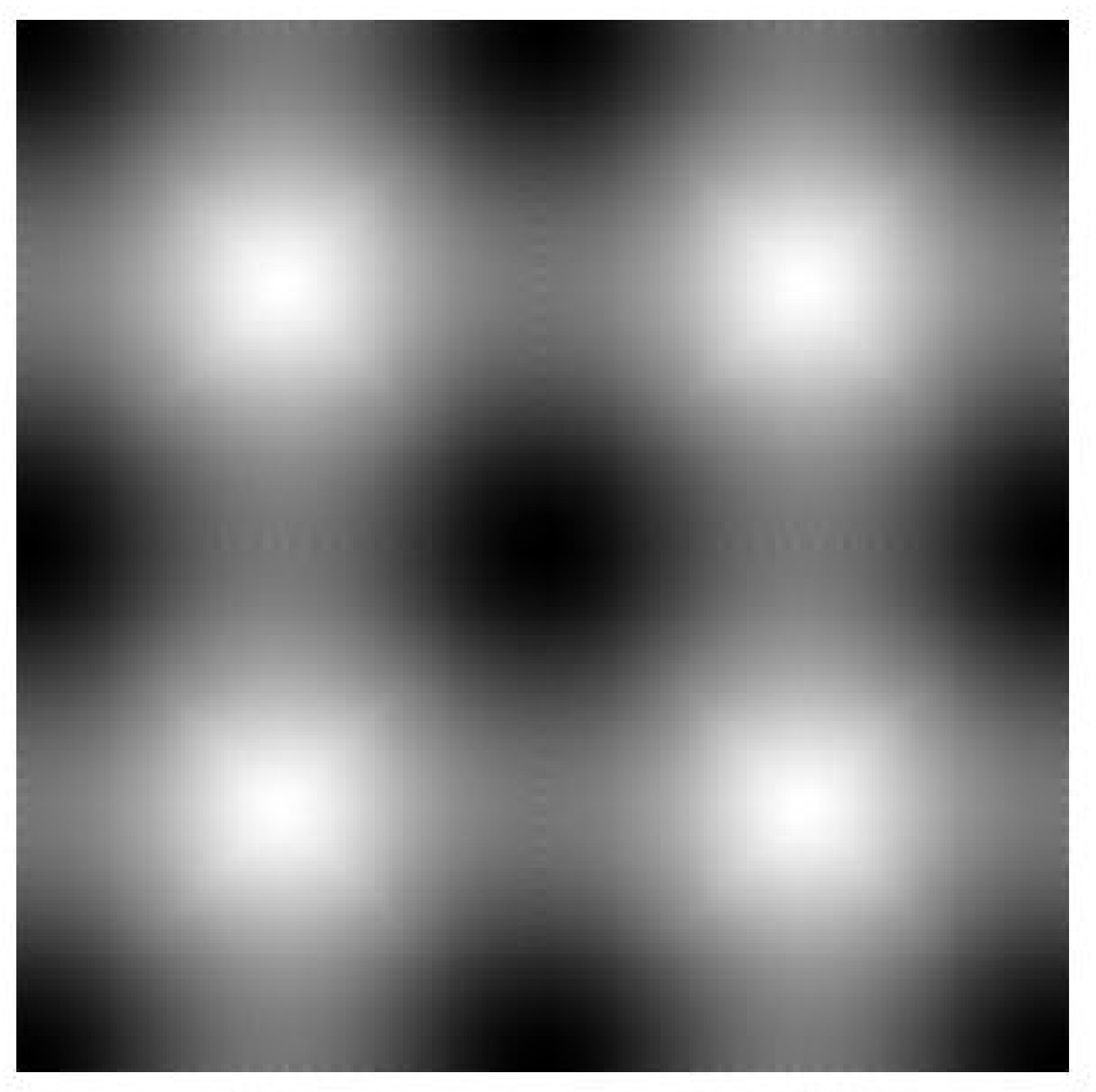}}\hfil
\epsfxsize=0.475\hsize\epsfbox{fig4678z.eps}\hspace*{-0.4225\hsize}%
\epsfxsize=0.370\hsize\raisebox{0.0517\hsize}{\epsfbox{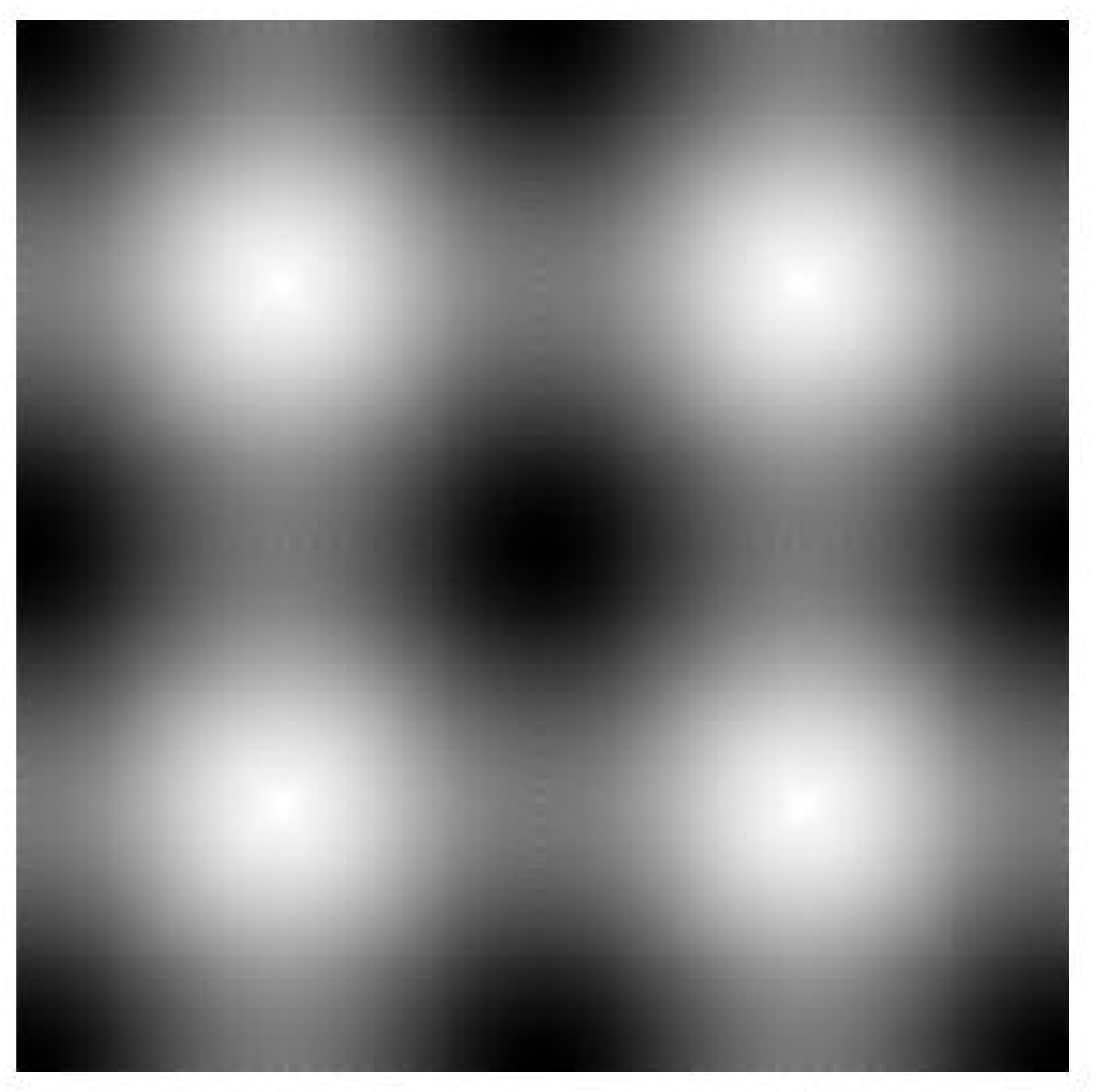}}
\hbox to\hsize{\hspace*{36pt}\scriptsize
(c) $j=2$ and $\alpha_2=(\sqrt{13}+3)/2$\hfil\hspace*{+24pt}
(d) $j=3$ and $\alpha_3=\sqrt 5+2$\hfil
\hspace*{-40pt}}\vskip0.2in
\caption{Fig.~4. Density plots of $1/\chi(q_x,q_y)$ for cases
when the sequences of rapidities ($u_m$) and ($v_m$) are
quasi-periodic sequences of two objects given by (\ref{abseq}) and
(\ref{rapid2}) at $k= 0.83791870$ ($\xi\approx 8$),
$T>T_{\rm c}$. There is no significant $j$-dependence.}
\end{figure}
 
To look at the situation more quantitatively, making sure that there are
indeed no incommensurate peaks, we can study $\chi(0,q)$ and
$\chi(q,q)$. We have plotted $\chi(q,q)$ versus $q$ for $j=0,\ldots,4$ and
$T<T_{\rm c}$ in Fig.~5$\,$(a), and also for $T>T_{\rm c}$ in
Fig.~5$\,$(b). As $j$ increases, there are more ${\rm B}$ type of
rapidity lines. This in turn means more weak bonds are present in the
system. Therefore, as $j$ increases, the peaks in the susceptibility
decrease, as shown in these plots. The changes are very small, however.
\begin{figure}[htbp]
\vskip-0.1in\hskip12pt\epsfclipon
\epsfxsize=0.40\hsize
\epsfbox{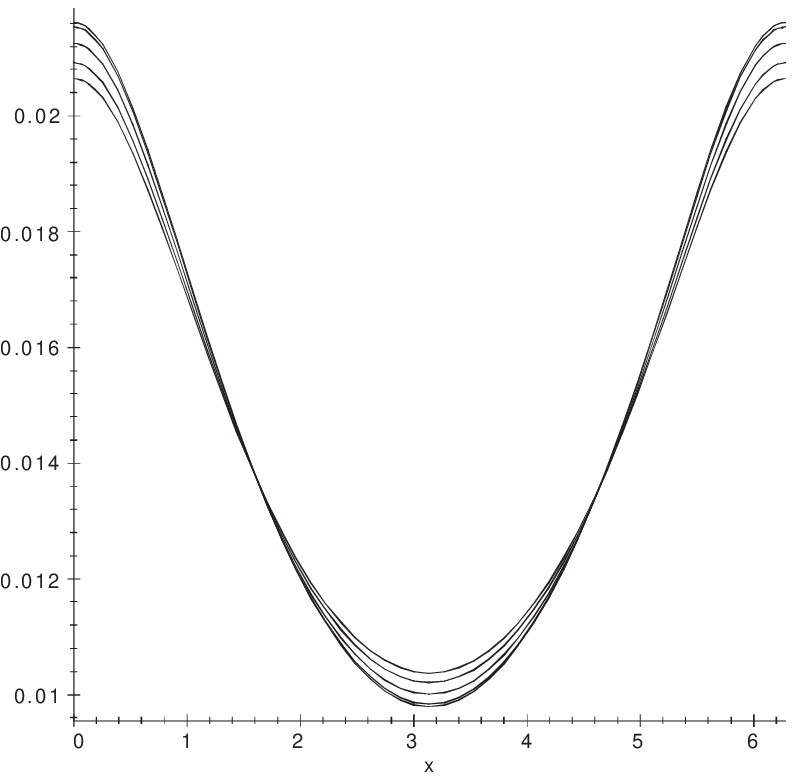}
\hskip40pt
\epsfxsize=0.40\hsize
\epsfbox{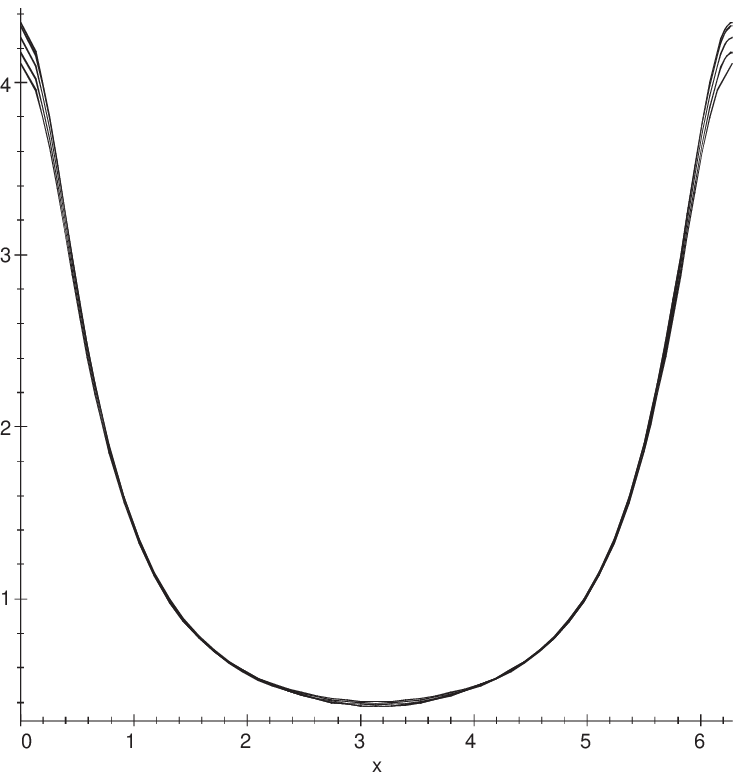}
\vskip 0.2in
\hbox to\hsize{\hspace*{48pt}\scriptsize
(a) $T<T_{\rm c}$: $k=4.2309029$\hfil
(b) $T>T_{\rm c}$: $k=0.2363561688$  ($\xi=1$)\hspace{-5pt}\hfil
\hspace*{-36pt}}
\vskip 0.2in
\caption{Fig.~5. Plots of $\chi(q,q)$ versus $\mathsf x\equiv q$ for the
cases given by (\ref{abseq}) and (\ref{rapid2}) and $j=0,\ldots,4$. The
curves for $j=0$ have the highest value at $q=0$, and the peaks decrease
in magnitude as $j$ increases.}
\end{figure}
The plots clearly show no indication of incommensurate peaks. The
behavior of $\chi({\bf q})$ for $T>T_{\rm c}$ is not much different from
the behavior at $T<T_{\rm c}$, except that the peaks are sharper.

We could give more density plots for different temperatures and also for
temperatures below and above $T_{\rm c}$. But those plots would not be
much different from Fig.~4. We find that as $T\to T_{\rm c}$, the peaks of
$\chi({\bf q})$ become sharper. Also, the peaks of $\chi({\bf q})$ for
$T>T_{\rm c}$ are sharper than those for $T<T_{\rm c}$, as the correlation
length above $T_{\rm c}$ is only half in length compared to the one at the
dual temperature below $T_{\rm c}$. \cite{MW} But it is hard to read that
off from a density plot.

\subsection{Sequences of Two Objects, Example II} 
Instead of (\ref{rapid2}), we may also choose
\begin{eqnarray}
u_{\rm B}=3\lambda/4,\quad u_{\rm A}=2\lambda/4,
\quad v_{\rm B}=\lambda/4,\quad v_{\rm A}=0.
\label{rapid2b}\end{eqnarray}
Comparing (\ref{bg}) and (\ref{gp}) with (\ref{g}) again, we find 
\begin{eqnarray}
\bg[m_2,m_3,m_0,m_1]=g[m_3,m_2,m_1,m_0],\nonumber\\
\quad g'[m_0,m_1,m_2,m_3]=g[m_3,m_2,m_1,m_0].
\end{eqnarray}
It is easily seen that (\ref{4fold}) still holds, so that
$\chi({\bf q})$ still has 4-fold  rotation symmetry. The behaviors of
$\chi({\bf q})$ are essentially the same as in the previous case, except
that the peaks become sharper as $j$ increases.

\subsection{Sequences of Two Objects, Example III} 
If we let
\begin{eqnarray}
u_{\rm A}=3\lambda/4,\quad u_{\rm B}=2\lambda/4,
\quad v_{\rm B}=\lambda/4,\quad v_{\rm A}=0,
\label{rapid2c}\end{eqnarray}
then
\begin{eqnarray}
\bg[m_3,m_2,m_0,m_1]=g[m_3,m_2,m_1,m_0],\nonumber\\
\quad g'[m_0,m_1,m_3,m_2]=g[m_3,m_2,m_1,m_0].
\end{eqnarray}
Consequently, (\ref{4fold}) no longer holds. As a result, $\chi({\bf q})$
behaves more like that of the rectangular Ising lattice, which is not
invariant under $90^{\circ}$ rotations, but still has only commensurate
peaks. Density plots are shown in Fig.~6.
\begin{figure}[htbp]
\vskip0.1in\hskip-2pt\epsfclipon
\epsfxsize=0.475\hsize\epsfbox{fig4678z.eps}\hspace*{-0.4225\hsize}%
\epsfxsize=0.370\hsize\raisebox{0.0517\hsize}{\epsfbox{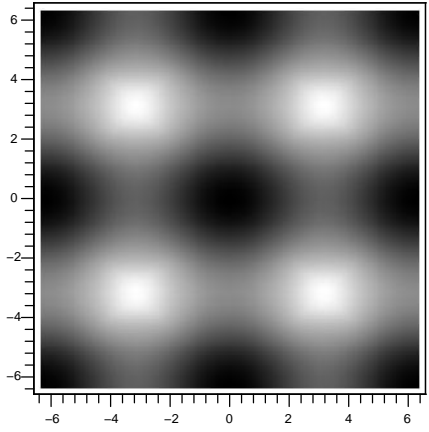}}\hfil
\epsfxsize=0.475\hsize\epsfbox{fig4678z.eps}\hspace*{-0.4225\hsize}%
\epsfxsize=0.370\hsize\raisebox{0.0517\hsize}{\epsfbox{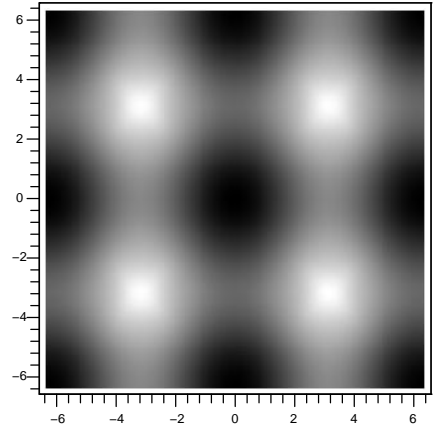}}
\hbox to\hsize{\hspace*{10pt}\scriptsize
(a) golden ratio: $j=0$ and $\alpha_0=(\sqrt 5+1)/2$\hfil
(b) silver mean: $j=1$ and $\alpha_1=\sqrt 2+1$\hfil
\hspace*{-48pt}}
\vskip0.3in\hskip0pt
\epsfxsize=0.475\hsize\epsfbox{fig4678z.eps}\hspace*{-0.4225\hsize}%
\epsfxsize=0.370\hsize\raisebox{0.0517\hsize}{\epsfbox{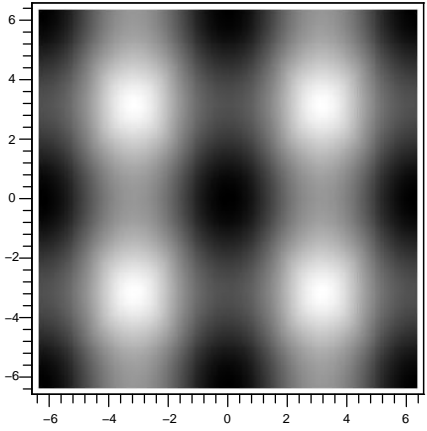}}\hfil
\epsfxsize=0.475\hsize\epsfbox{fig4678z.eps}\hspace*{-0.4225\hsize}%
\epsfxsize=0.370\hsize\raisebox{0.0517\hsize}{\epsfbox{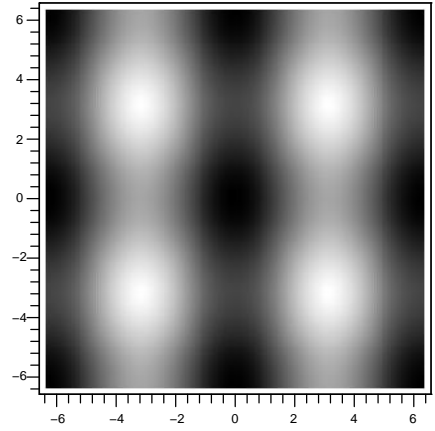}}
\hbox to\hsize{\hspace*{35pt}\scriptsize
(c) $j=2$ and $\alpha_2=(\sqrt{13}+3)/2$\hfil\hspace*{48pt}
(d) $j=3$ and $\alpha_3=\sqrt 5+2$\hfil
\hspace*{-16pt}}
\vskip0.2in
\caption{Fig.~6. Density plots of $1/\chi(q_x,q_y)$ for the cases defined
by (\ref{abseq}) and (\ref{rapid2c}) at $T>T_{\rm c}$, $k=0.49127583$
($\xi\approx 2$). The susceptibility is like the one of the rectangular
Ising model, as can be seen with some effort, with peaks still at the
commensurate positions.}
\end{figure}

\subsection{Sequences of Three Objects, Example IV} 
We now let the sequence of rapidity lines be defined by (\ref{abcseq})
and let 
\begin{eqnarray}
u_{\rm A}=3\lambda/4,\quad u_{\rm B}=2\lambda/4,
\quad u_{\rm C}=\lambda/4,\quad v=0.
\label{rapid3}\end{eqnarray}
Comparing (\ref{hg}) and (\ref{hgp}) with (\ref{g}), we obtain
\begin{eqnarray}
\hg[m_3,m_2,m_1,m_0]=g[m_3,m_2,m_1,m_0],\nonumber\\
\quad \hg'[m_0,m_3,m_2,m_1]=g[m_3,m_2,m_1,m_0].
\label{hgg} \end{eqnarray}
We evaluate the $\chi({\bf q})$ in (\ref{chis}) by substituting this
equation into (\ref{avcor3a}), (\ref{avcor3b}), (\ref{avcor3c}),
(\ref{avcor3aa}), (\ref{avcor3bb}) and (\ref{avcor3cc}). The
probabilities for the three-object sequence given by (\ref{prob1})
through (\ref{prob3}) are quite complicated. Nevertheless, we find
similar behavior for all different $j$'s and temperatures.
There is no incommensurate behavior---the peak of the susceptibility
$\chi({\bf q})$ is at the commensurate position of the ordinary Ising
model, $(q_x,q_y)=(0,0)$, and repeated periodically with periods $2\pi$.

In Fig.~7, four density plots are presented for $T<T_{\rm c}$ at
$k=1.1934332$ and for $j=1,\ldots,4$. We again find that $\chi({\bf q})$
decreases as $j$ increases. Since only the ($u_m$) sequence is aperiodic,
the distortion due to the quasiperiodicity on
$\chi({\bf q})$ is along the diagonal. In this particular case, we find
that the two diagonals are the symmetry axes of $\chi({\bf q})$. 
\begin{figure}[htbp]
\vskip0.1in\hskip-2pt\epsfclipon
\epsfxsize=0.475\hsize\epsfbox{fig4678z.eps}\hspace*{-0.4225\hsize}%
\epsfxsize=0.370\hsize\raisebox{0.0517\hsize}{\epsfbox{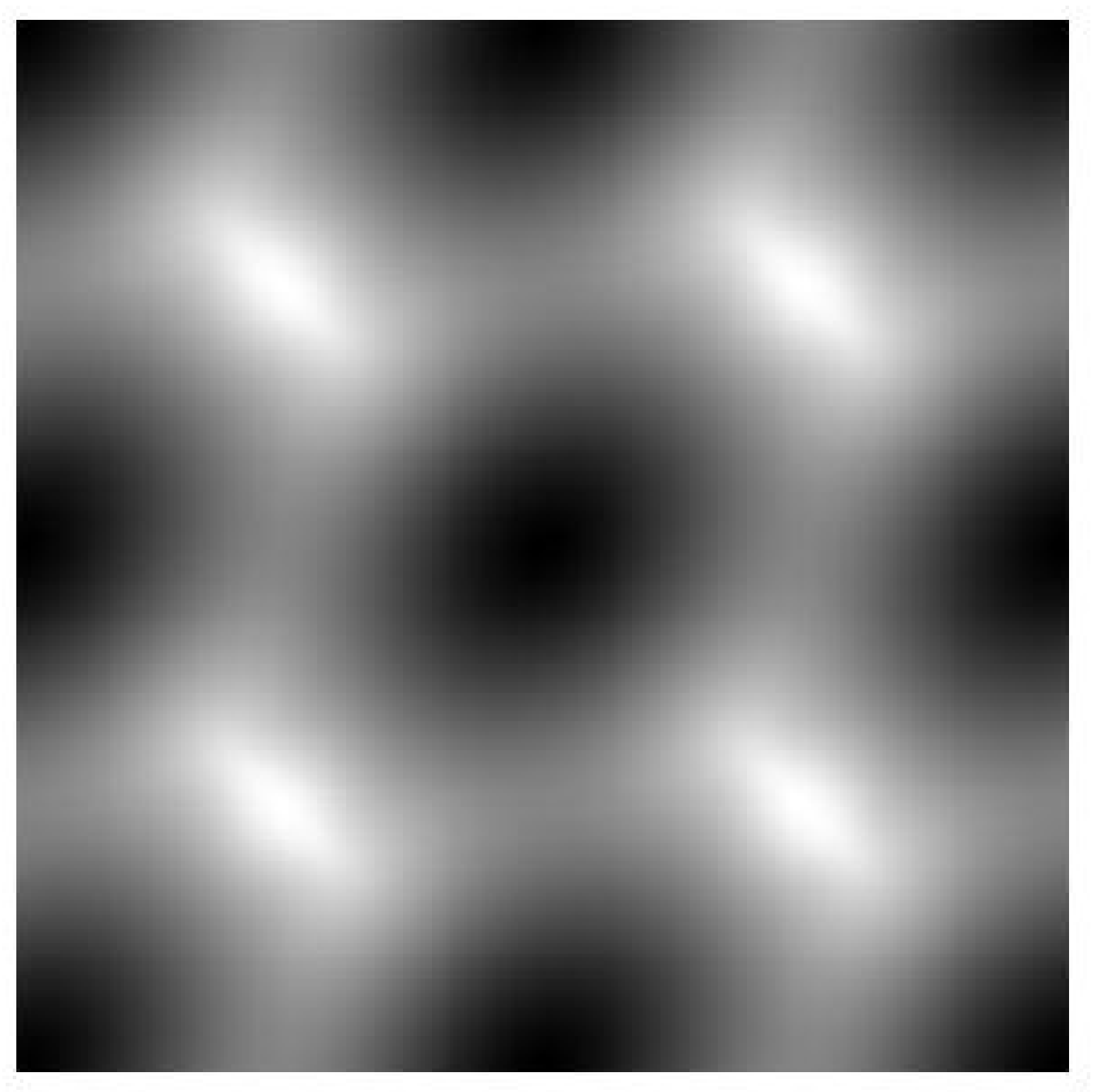}}\hfil
\epsfxsize=0.475\hsize\epsfbox{fig4678z.eps}\hspace*{-0.4225\hsize}%
\epsfxsize=0.370\hsize\raisebox{0.0517\hsize}{\epsfbox{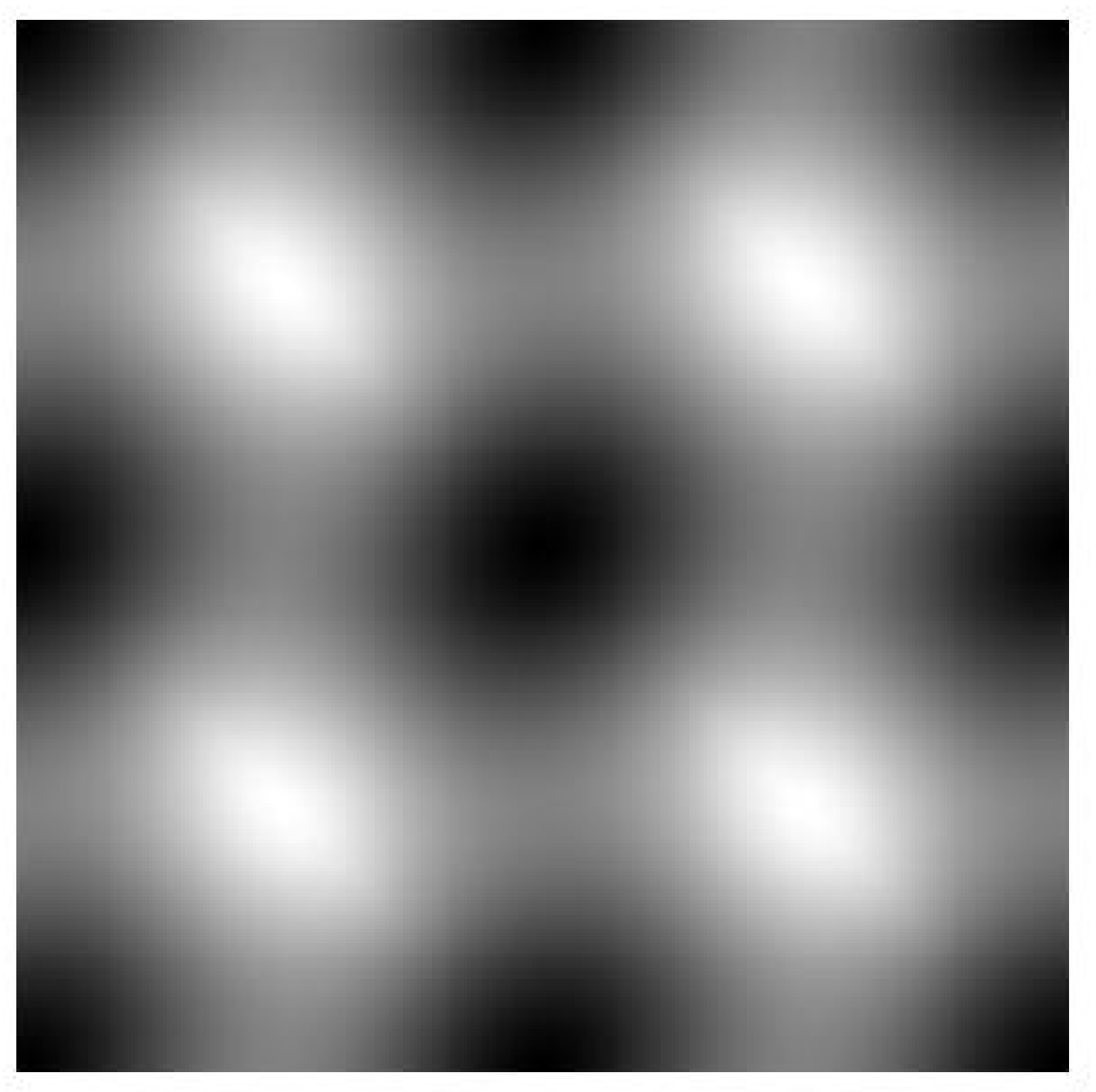}}
\hbox to\hsize{\hspace*{18pt}\scriptsize
(a) silver mean: $j=1$ and $\alpha_1=\sqrt 2+1$\hfil
(b) $j=2$ and $\alpha_2=(\sqrt{13}+3)/2$\hfil
\hspace*{-48pt}}
\vskip0.2in\hskip0pt
\epsfxsize=0.475\hsize\epsfbox{fig4678z.eps}\hspace*{-0.4225\hsize}%
\epsfxsize=0.370\hsize\raisebox{0.0517\hsize}{\epsfbox{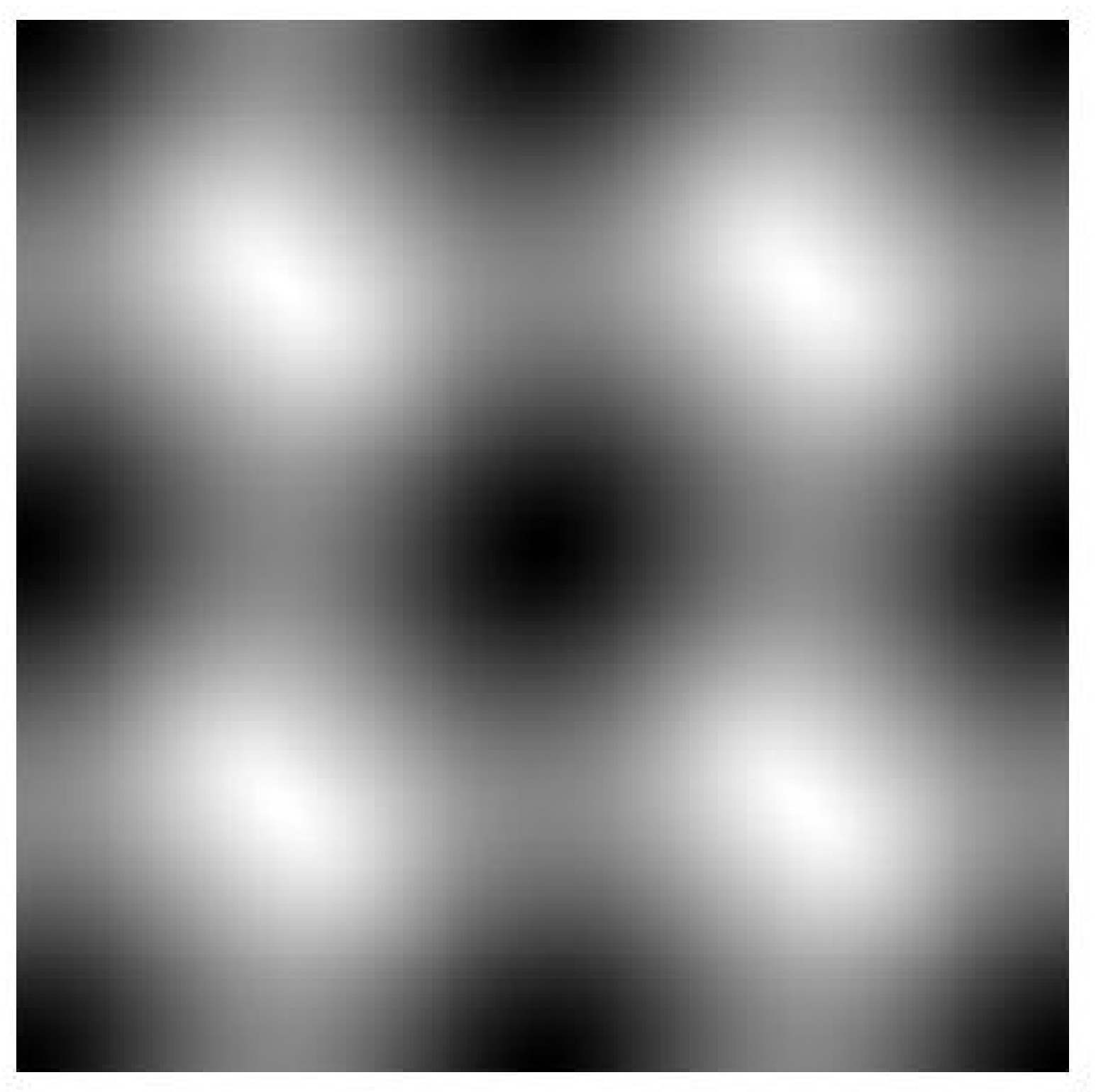}}\hfil
\epsfxsize=0.475\hsize\epsfbox{fig4678z.eps}\hspace*{-0.4225\hsize}%
\epsfxsize=0.370\hsize\raisebox{0.0517\hsize}{\epsfbox{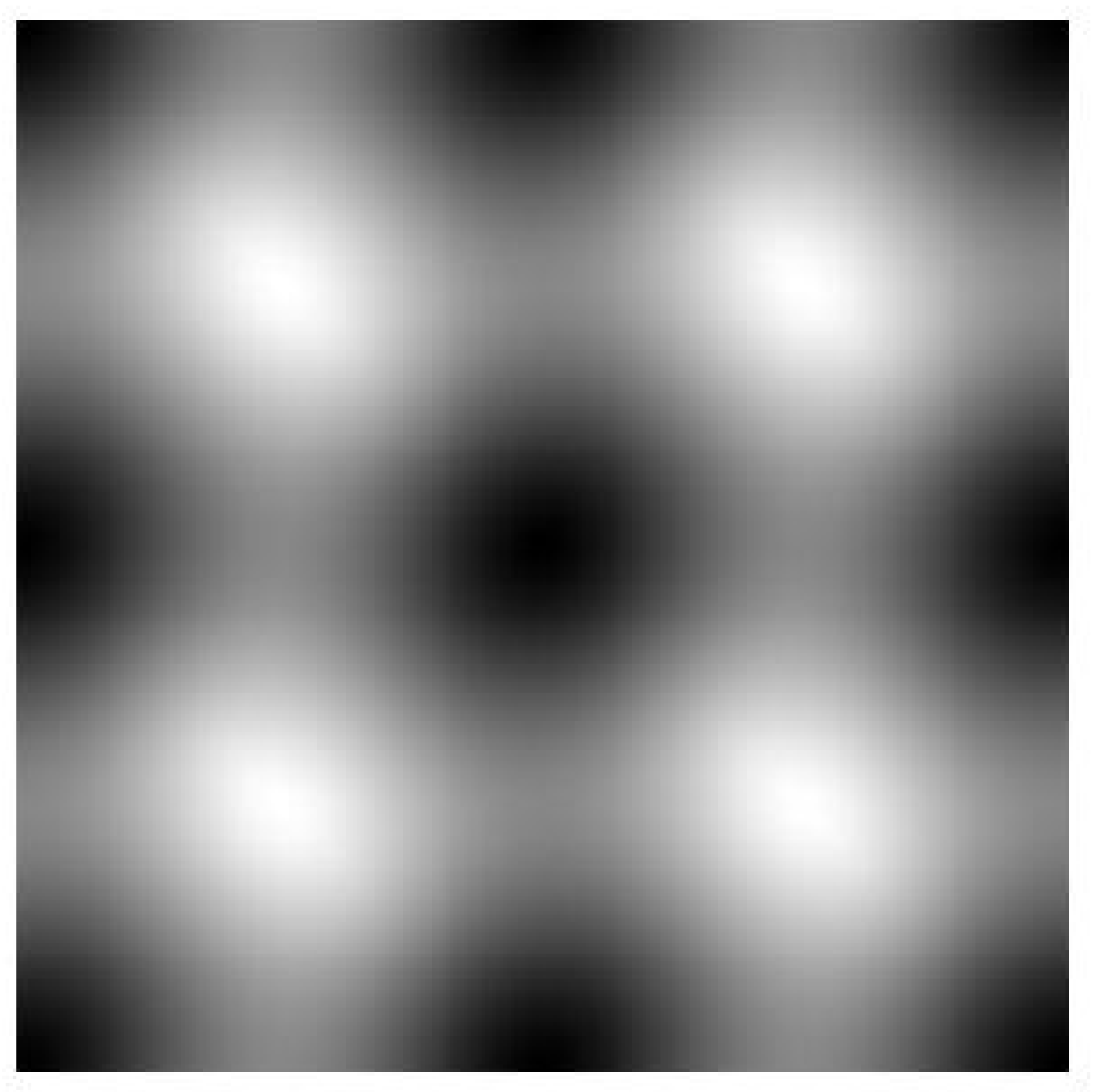}}
\hbox to\hsize{\hspace*{46pt}\scriptsize
(c) $j=3$ and $\alpha_3=\sqrt 5+2$\hfil\hspace*{60pt}
(d) $j=4$ and $\alpha_4=(\sqrt{29}+5)/2$\hfil
\hspace*{-12pt}}
\vskip0.2in
\caption{Fig.~7. Density plots of $1/\chi(q_x,q_y)$ for cases when only
($u_m$) is a quasi-periodic sequence. They are given by (\ref{abcseq})
and (\ref{rapid3}) at $T<T_{\rm c}$, $k=1.1934332$. Again
the peaks are only at the commensurate positions. They are elongated
in a diagonal direction.}
\end{figure}

\subsection{Sequences of Three Objects, Example V} 
If instead of (\ref{rapid3}), we let
\begin{eqnarray}
u_{\rm A}=3\lambda/4,\quad u_{\rm C}=2\lambda/4,
\quad u_{\rm B}=\lambda/4,\quad v=0,
\label{rapid3b}\end{eqnarray}
then
\begin{eqnarray}
\hg[m_3,m_1,m_2,m_0]=g[m_3,m_2,m_1,m_0],\nonumber\\
\quad \hg'[m_0,m_3,m_1,m_2]=g[m_3,m_2,m_1,m_0].
\label{hagg} \end{eqnarray}
The resulting $q$-dependent susceptibility is less symmetric. Four
density plots at $T<T_{\rm c}$, $k=1.1934332$, are shown in Fig.~8 for
$j=1,\ldots,4$. Again we only find commensurate peaks.
\begin{figure}[htbp]
\vskip0.1in\hskip-2pt\epsfclipon
\epsfxsize=0.475\hsize\epsfbox{fig4678z.eps}\hspace*{-0.4225\hsize}%
\epsfxsize=0.370\hsize\raisebox{0.0517\hsize}{\epsfbox{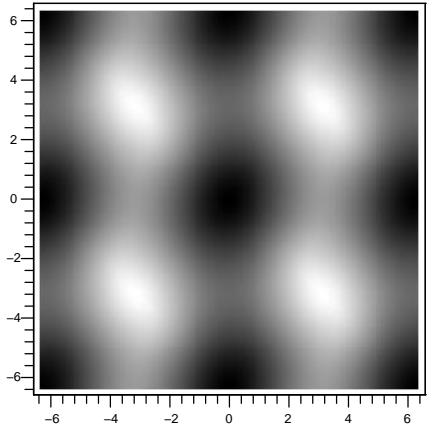}}\hfil
\epsfxsize=0.475\hsize\epsfbox{fig4678z.eps}\hspace*{-0.4225\hsize}%
\epsfxsize=0.370\hsize\raisebox{0.0517\hsize}{\epsfbox{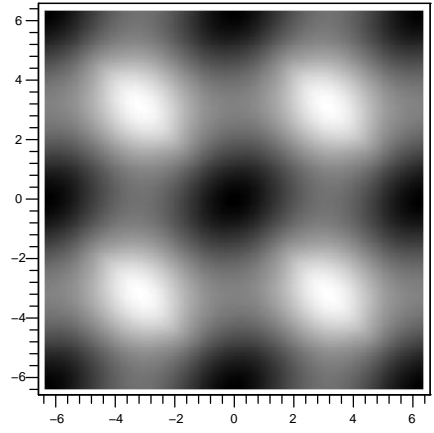}}
\hbox to\hsize{\hspace*{18pt}\scriptsize
(a) silver mean: $j=1$ and $\alpha_1=\sqrt 2+1$\hfil
(b) $j=2$ and $\alpha_2=(\sqrt{13}+3)/2$\hfil
\hspace*{-48pt}}
\vskip0.2in\hskip0pt
\epsfxsize=0.475\hsize\epsfbox{fig4678z.eps}\hspace*{-0.4225\hsize}%
\epsfxsize=0.370\hsize\raisebox{0.0517\hsize}{\epsfbox{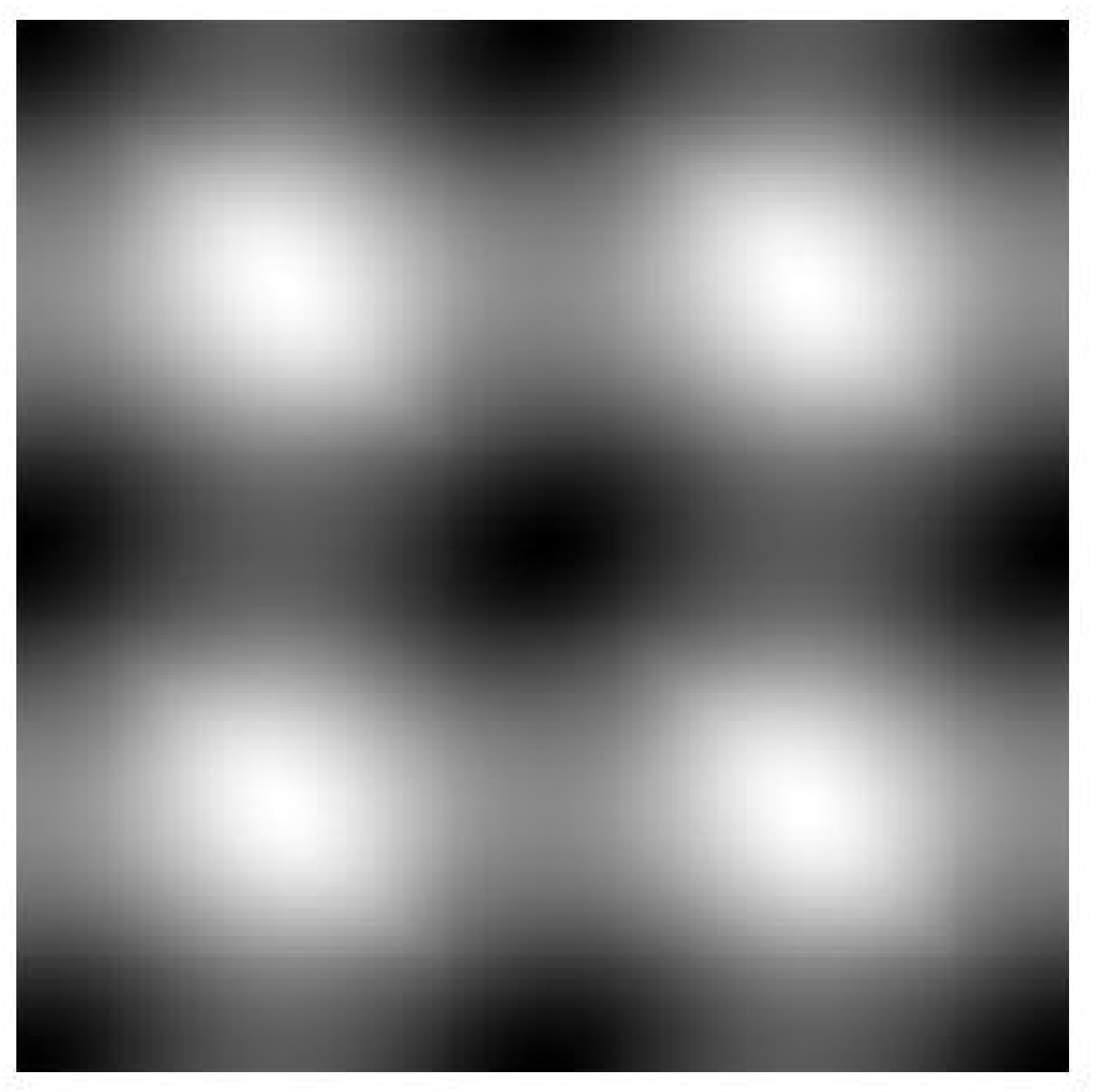}}\hfil
\epsfxsize=0.475\hsize\epsfbox{fig4678z.eps}\hspace*{-0.4225\hsize}%
\epsfxsize=0.370\hsize\raisebox{0.0517\hsize}{\epsfbox{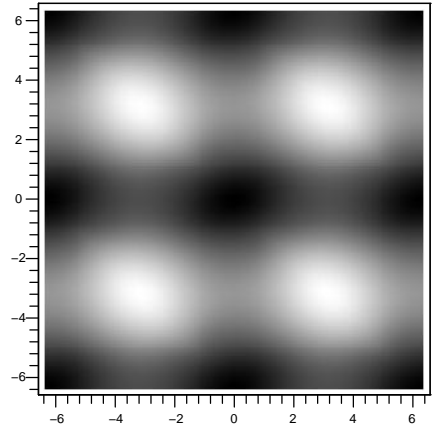}}
\hbox to\hsize{\hspace*{46pt}\scriptsize
(c) $j=3$ and $\alpha_3=\sqrt 5+2$\hfil\hspace*{52pt}
(d) $j=4$ and $\alpha_4=(\sqrt{29}+5)/2$\hfil
\hspace*{-18pt}}
\vskip0.2in
\caption{Fig.~8. Density plots of $1/\chi(q_x,q_y)$ for the cases given by
(\ref{abcseq}) and (\ref{rapid3b}) at $T<T_{\rm c}$, $k=1.1934332$.
Still the peaks of $\chi(q_x,q_y)$ are at the commensurate positions.
The peaks are now elongated and a slight dependence on $j$ may be
observed.}
\end{figure}

\section{A Mixed Case}

We have examined quasiperiodic Ising lattices on a square lattice, whose
interactions are quasiperiodic and ferromagnetic, and we have found very
similar commensurate behaviors.

Things change dramatically if we consider mixed cases with both ferro-
and antiferromagnetic interactions, as we already know from our previous
work that there will be many incommensurate peaks within the unit cell as
the temperature moves close to the critical value. \cite{AJPq,APmc1}
There is one new aspect: The results, especially the positions of the
many incommensurate peaks, are heavily dependent on the value of $j$. We
shall illustrate this with one example based on some ideas of Section
5 of Ref.\ \citen{AJPq}, where several $j=0$ cases have been studied.

Unlike the ferromagnetic case, we can now construct an example starting
from the symmetric square-lattice Ising model and flipping the signs
of the couplings by site-dependent gauge transformations. Using
Theorem 1, Eqs.\ (5.17) and (5.18) of Ref.\ \citen{AJPq} are now
replaced by
\begin{eqnarray}
\phi^{(j)}(m)&=&
(-1)^{\lfloor m/\alpha_j\rfloor}(1-2\{m/\alpha_j\}),\nonumber\\
&=&\sum_{l=-\infty}^{\infty} 
\frac{{\rm e}^{2\pi{\rm i}(l+1/2)m/\alpha_j}}{(l+1/2)^2\pi^2}=
\phi^{(j)}(-m).
\label{2.13}
\end{eqnarray}
Choosing a model aperiodic in both diagonal directions as in Section 5.6
of Ref.\ \citen{AJPq}, the averaged connected correlation function now
becomes
\begin{equation}
C^{\rm(c)}(l,m)=\phi^{(j)}(l+m)\phi^{(j)}(l-m)C_0^{\rm(c)}(l,m),
\label{2.21}
\end{equation}
with $C_0^{\rm(c)}(l,m)$ the connected pair-correlation function of
the square-lattice Ising model. This implies that $\chi({\bf q})$
has many incommensurate peaks within the unit cell, and is given by 
\begin{equation}
\chi(q_x,q_y)=
\sum_{l=-\infty}^{\infty}\sum_{m=-\infty}^{\infty}\frac
{\chi^{~}_0\Big(q_x+2\pi(l+m+1)/\alpha_j,q_y+2\pi(l-m)/\alpha_j\Big)}
{(l+1/2)^2(m+1/2)^2\pi^4},
\label{2.16dd}
\end{equation}
with $\chi^{~}_0({\bf q})$ the wavevector-dependent susceptibility of
the regular square-lattice Ising model.

\begin{figure}[htbp]
\vskip0.1in\hskip-2pt\epsfclipon
\epsfxsize=0.475\hsize\epsfbox{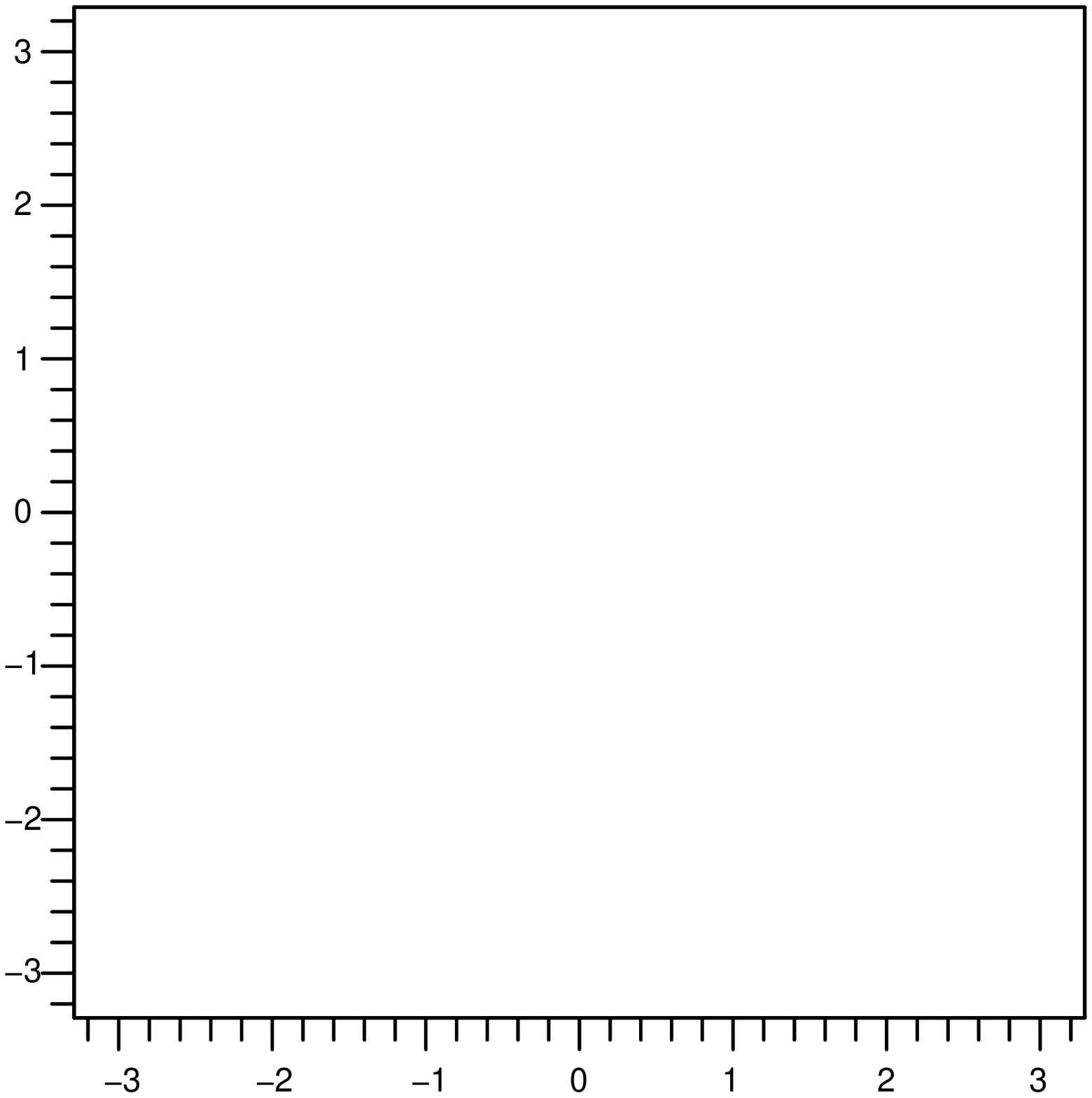}\hspace*{-0.4225\hsize}%
\epsfxsize=0.370\hsize\raisebox{0.0517\hsize}{\epsfbox{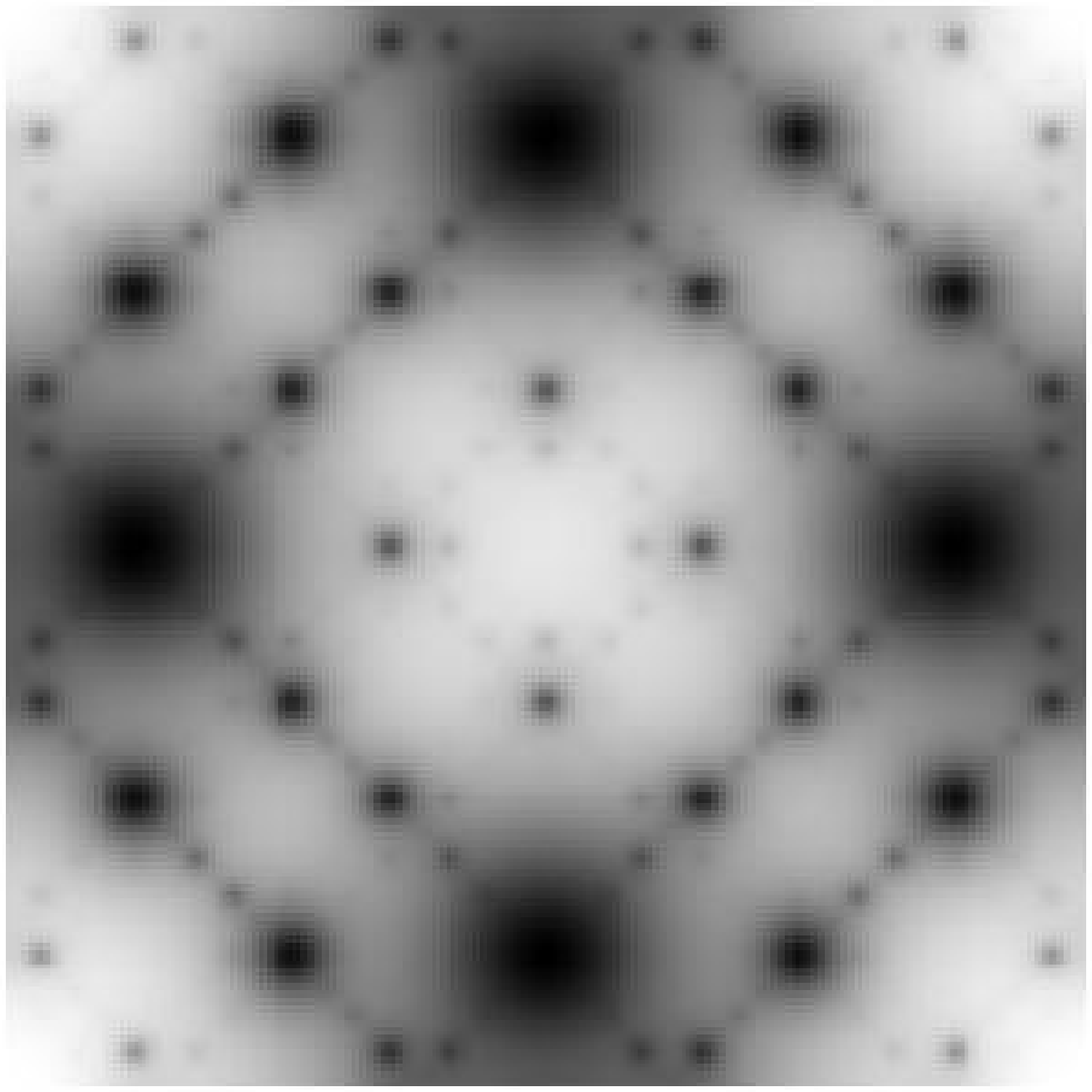}}\hfil
\epsfxsize=0.475\hsize\epsfbox{fig9z.eps}\hspace*{-0.4225\hsize}%
\epsfxsize=0.370\hsize\raisebox{0.0517\hsize}{\epsfbox{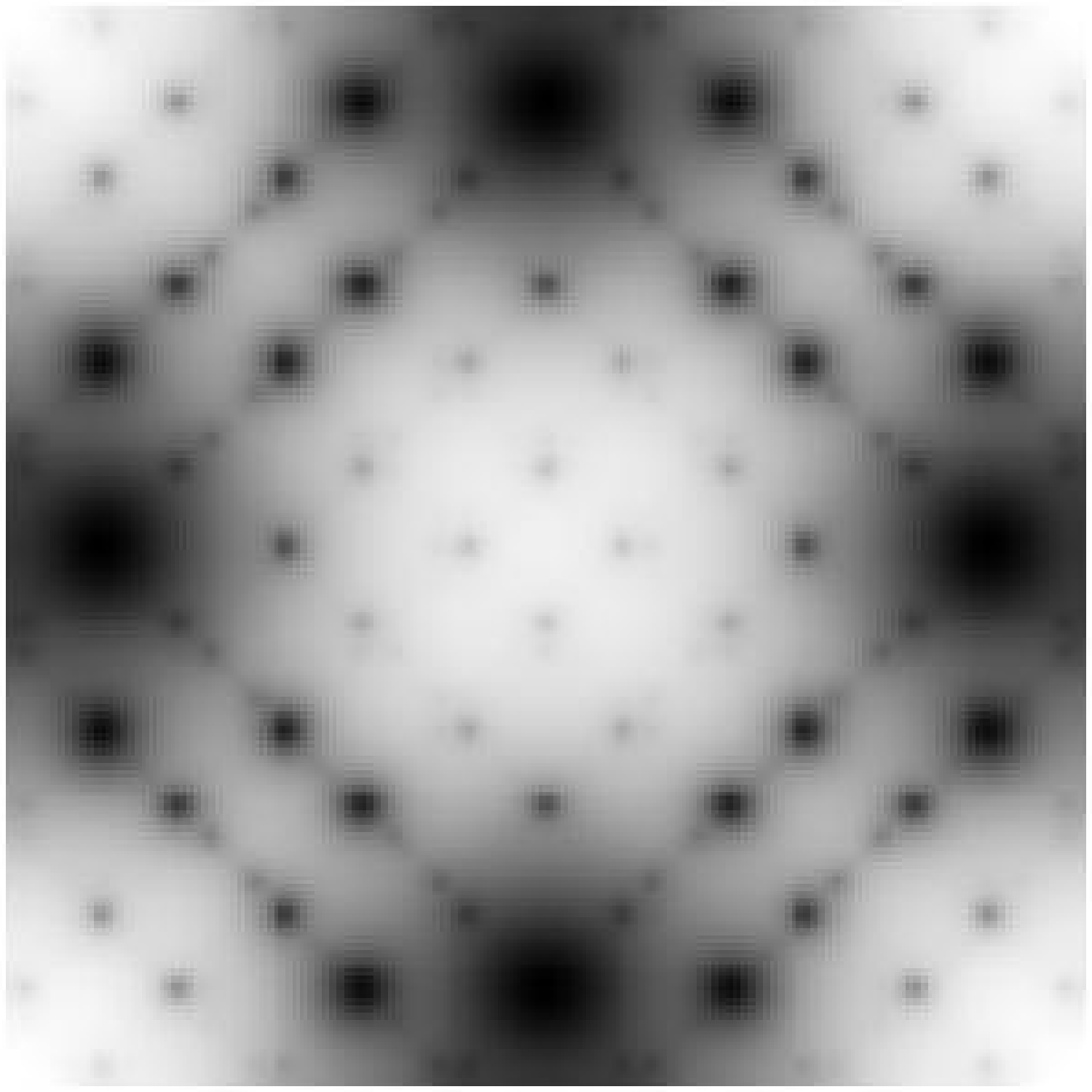}}
\hbox to\hsize{\hspace*{20pt}\scriptsize
(a) golden ratio: $j=0$, $\alpha_1=(\sqrt5+1)/2$\hfil\hspace*{5pt}
(b) silver mean: $j=1$, $\alpha_1=\sqrt2+1$\hspace*{9pt}\hfil
\hspace*{-30pt}}
\vskip0.1in\hskip-2pt
\epsfxsize=0.475\hsize\epsfbox{fig9z.eps}\hspace*{-0.4225\hsize}%
\epsfxsize=0.370\hsize\raisebox{0.0517\hsize}{\epsfbox{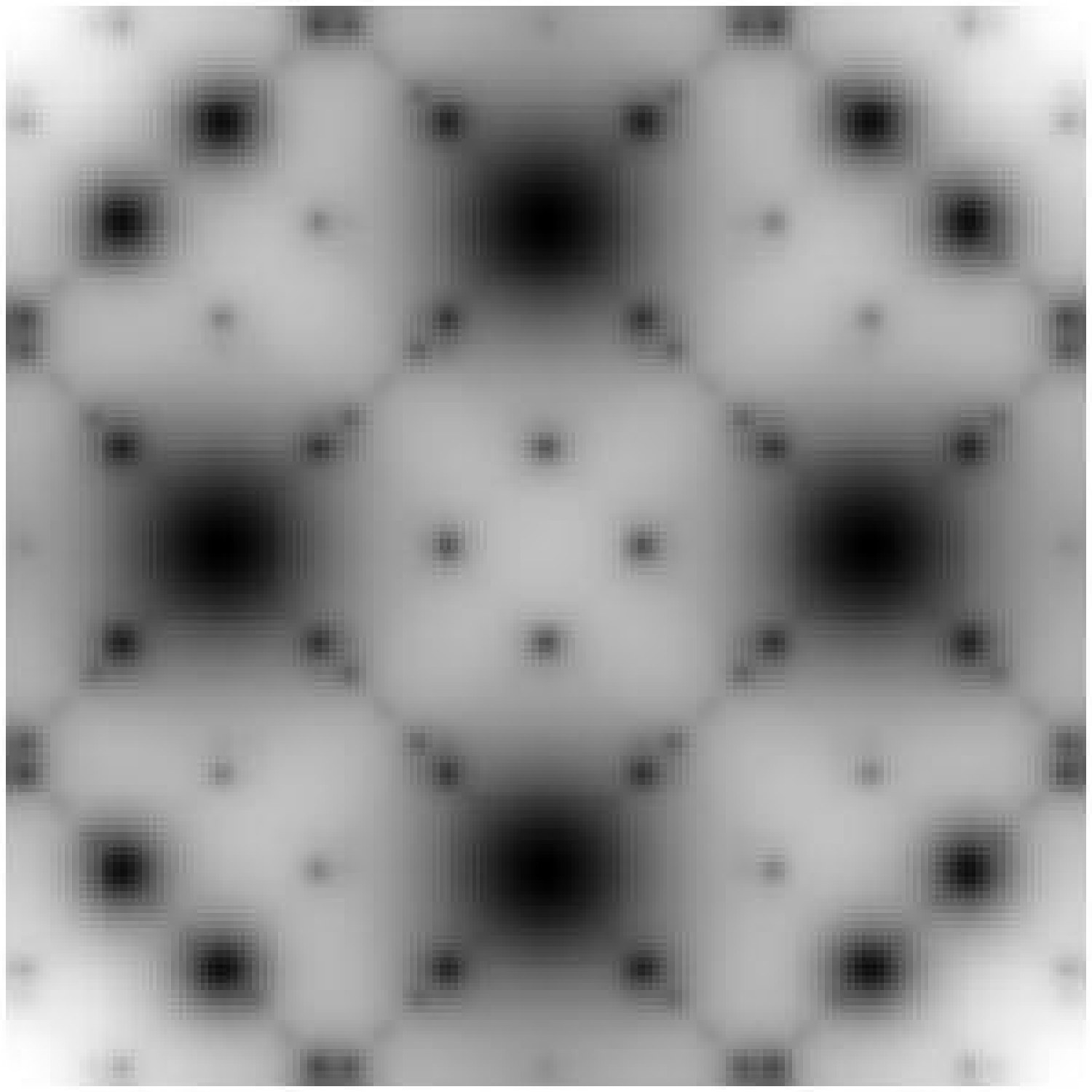}}\hfil
\epsfxsize=0.475\hsize\epsfbox{fig9z.eps}\hspace*{-0.4225\hsize}%
\epsfxsize=0.370\hsize\raisebox{0.0517\hsize}{\epsfbox{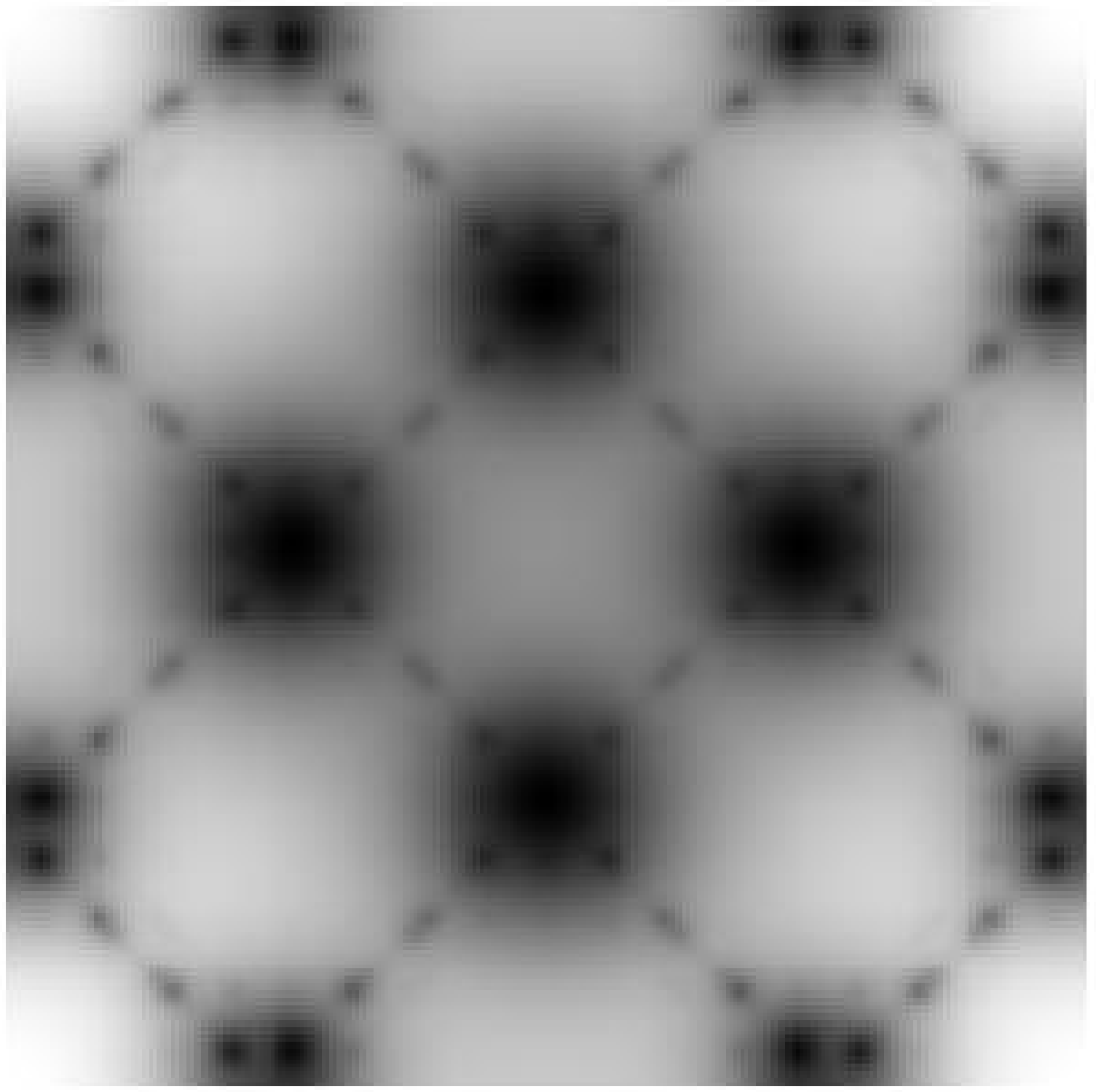}}
\hbox to\hsize{\hspace*{36pt}\scriptsize
(c) $j=2$, i.e.\ $\alpha_1=(\sqrt13+3)/2$\hfil\hspace*{26pt}
(d) $j=3$, i.e.\ $\alpha_3=\sqrt5+2$\hfil
\hspace*{-16pt}}
\vskip0.2in
\caption{Fig.~9. Density plots of $1/\chi(q_x,q_y)$ for the mixed case
for four values of $j$, ($j=0,\ldots,3$), with $q_x$ and $q_y$ in the
interval $(-\pi,\pi)$ and $k=0.915398728\cdots$. Now there are many
incommensurate peaks and their positions depend strongly on $j$.
The principal peaks are at ($\pm q_j,0)$, ($0,\pm q_j)$, with
$q_0=2\pi(1-1/\alpha_0)=2.39996\cdots$,
$q_1=2\pi/\alpha_1=2.60258\cdots$, $q_2=2\pi/\alpha_2=1.90239\cdots$,
$q_3=2\pi/\alpha_3=1.48325\cdots$. This last value is close to $\pi/2$,
which is reflected in figure (d).}
\end{figure}
Density plots are given in Fig.~9 for four cases with correlation length
$\xi\approx16$. Clearly, the results depend strongly on $j$. The case
$j=3$ is the most different as it almost looks like the periods have
been halved. This can be explained easily since
$2\pi/\alpha_3\approx\pi/2$.

\section{Conclusions}
From the current work and our previous papers \cite{APpenta,AJPq,APmc1}
we can draw several conclusions:
\begin{itemize}
\item
The wavevector-dependent susceptibilities $\chi({\bf q})$ of models,
whose spin sites are on regular lattices, are always periodic. This
includes cases when the interactions between the spins are
quasi-periodic.
\item
When the interactions between spins are quasiperiodic, but strictly
ferromagnetic, $\chi({\bf q})$ has only commensurate peaks, with behavior
very similar to that of the regular Ising model.
\item
The $q$-dependent susceptibilities $\chi({\bf q})$ of models on regular
periodic lattices can have everywhere-dense incommensurate peaks in every
unit cell, but only for cases for which the interactions between spins
are mixed with both ferro- and antiferromagnetic couplings present.
\cite{AJPq, APmc1}
\item
When the lattice is quasiperiodic---such as a $Z$-invariant Ising model
on Penrose tiles---$\chi({\bf q})$ is no longer periodic but
quasiperiodic and exhibits everywhere-dense incommensurate peaks, even
for the case of purely ferromagnetic couplings. Only few of these peaks
are visible within a given limited area of ${\bf q}$-space when the
temperature is far away from the critical temperature. The number of
visible peaks increases as $T$ approaches $T_{\rm c}$. \cite{APpenta}
\end{itemize}
There are many other quasiperiodic sequences. Still we have examined
a variety of cases and believe that the above conclusions are quite
generic. 

It may be interesting to consider the $q$-dependent susceptibility
$\chi({\bf q})$ of the $Z$-invariant Ising model on the labyrinth
\cite{BGB, GBS, GB} for which the distances between the spins are
also aperiodic. To obey the symmetry, the couplings of pairs of spins
must be related to the distances between the spins. When the distances
are equal, the corresponding couplings must be chosen to be equal. Since
the coupling $K$ and ${\bar K}$ in a $Z$-invariant model are related by
(\ref{couplings}), our preliminary efforts in this regard have not been
successful, but the model deserves further investigation. One thing we
can predict: The $q$-dependent susceptibility $\chi({\bf q})$, for the
Ising model on the labyrinth, can no longer be periodic, and its peaks
should be at incommensurate positions.

\section*{\sffamily\bfseries\normalsize ACKNOWLEDGMENTS}
\par\hspace*{\parindent}%
We are most thankful to  Dr.\ M.\ Widom for his interest in using exactly
solvable models to study quasicrystals. This work has been supported by
NSF Grant PHY 01-00041.

\catcode`\@=11
\def\footnotesize{\@setsize\footnotesize{12pt}\xpt\@xpt
\abovedisplayskip 10pt plus2pt minus5pt\belowdisplayskip \abovedisplayskip
\abovedisplayshortskip
\z@ plus3pt\belowdisplayshortskip 6pt plus3pt minus3pt
\def\@listi{\leftmargin\leftmargini
\topsep 6pt plus 2pt minus 2pt\parsep 0pt
plus 0pt minus 0pt
\itemsep \parsep}}
\catcode`\@=12
\footnotesize

\end{document}